\documentclass[a4paper,12pt,reqno,superscriptaddress,showkeys,nofootinbib]{revtex4}
\usepackage[centertags]{amsmath}
\usepackage{amsfonts}
\usepackage{amssymb}
\usepackage{amsthm}
\usepackage{newlfont}
\usepackage{stmaryrd}
\usepackage{mathrsfs}
\usepackage{mathtools}
\usepackage{euscript}
\usepackage{graphicx}
\usepackage{enumerate}
\usepackage[normalem]{ulem} % for strikeout text with \sout

\usepackage{color}
\definecolor{olive-green}{RGB}{60, 128, 49}

\usepackage{tikz}
\usepackage{pgf}
\usetikzlibrary{positioning,fit,calc}
\usetikzlibrary{automata,shapes}
\usetikzlibrary {arrows.meta}
\usepackage{wrapfig}
\usepackage{subfigure}
\usepackage{amscd}
\usepackage{hyperref}

\usepackage{changes}

\theoremstyle{plain}

\theoremstyle{definition}

\theoremstyle{remark}

\newcommand{\be}{\begin{equation}}
\newcommand{\en}{\end{equation}}

\newcommand{\opunit}{\text{1}\kern-0.22em\text{l}}

\newcommand{\id}{\textrm{d}}

\DeclareMathAlphabet{\mathpzc}{OT1}{pzc}{m}{it}

\let\oldsqrt\sqrt
\def\sqrt{\mathpalette\DHLhksqrt}
\def\DHLhksqrt#1#2{%
	\setbox0=\hbox{$#1\oldsqrt{#2\,}$}\dimen0=\ht0
	\advance\dimen0-0.2\ht0
	\setbox2=\hbox{\vrule height\ht0 depth -\dimen0}%
	{\box0\lower0.4pt\box2}}

\let\be=\beta

\DeclareMathAlphabet{\mathpzc}{OT1}{pzc}{m}{it}

\def\bea{\begin{eqnarray}}
\def\eea{\end{eqnarray}}
\def\ba{\begin{array}}
	\def\ea{\end{array}}

\usepackage{changes}

\begin{document}

\title{Frenetic steering: nonequilibrium-enabled navigation}
\author{Bram Lefebvre and Christian Maes\\
Department of Physics and Astronomy, KU Leuven, B-3001 Leuven, Belgium}
\email{christian.maes@kuleuven.be}%\affiliation{Gent, Belgium}
%\author{Christian Maes}
%\affiliation{Instituut voor Theoretische Fysica, KU Leuven, Belgium}

%\date{\today}

\begin{abstract}

%We study the possibility to navigate a probe coupled to a nonequilibrium medium by using time-symmetric parameters as control.

We explain the  steering of slow degrees of freedom by coupling them to driven components for which the time-symmetric reactivities are manipulated. We present the strategy and main principle that make  that sort of navigation feasible. For illustration, nonlinear limit cycles (as in the van der Pol oscillator) and strange attractors (as in the Lorenz dynamics) are seen to emerge when the driving in the nonequilibrium medium is kept fixed while the frenesy is tuned to produce the required forces.  We imagine that such frenetic control is available in Life as well, allowing selection of the appropriate biological functioning.
   
%it possible to provide any force by appropriately choosing the time-symmetric parameters. This means that 

%In particular, we simulate trajectories or a dynamics of slow variables from manipulating a {\it hidden} nonequilibrium system coupled to the slow variables. That navigation is possible by a {\it twist} between the path-dependent entropy production and frenesy.
\end{abstract}

\maketitle

%\tableofcontents
\section*{Significance}
{\bf The paper launches a new idea for the steering of probes (slow particles).  In many applications steering is done by deriving the force from a potential landscape.  Here we couple the probe to a nonequilibrium (driven or active) medium, and we control the frenesy, which means manipulating time-symmetric reactivities in the nonequilibrium medium.  That creates or alters the mean force on the probe, for producing the wished-for trajectories.   The algorithm is illustrated by steering a probe to follow the trajectories of a van der Pol oscillator and of the Lorenz system.   That novel scheme has both fundamental and applied significance, as it enables to deliver controlled motion in a nonequilibrium background.}

\section{Introduction}

An old example of steering is encountered in irrigation networks. In the case of the Sierra Nevada, more than 3,000 km of irrigation channels were built on its slopes by the settlers of the Umayyad conquest of Hispania (8th century), \cite{islamic_gardens}. The water running downhill is ingeniously diverted to flow along the different cultivations.  Such controlled use of a potential (gravity in the case of surface irrigation) combined with landscaping takes a more abstract formulation in gradient flow, where descent follows a free energy profile and a local measure of distance (or metric) determines what is steepest, obtaining the gradient descent method as first proposed by Cauchy (1847), \cite{cauchy}.\\

In the case of individual particles where a specific probe or collective variable is to follow a given trajectory, steering requires the more or less direct creation or alteration of the force on that probe. One way is to derive that force from a potential landscape or from a time-dependent external field.  However, when the probe is coupled to a nonequilibrium medium, a newer fascinating possibility arises, the subject of the present paper.\\
It has indeed been realized before that violating detailed balance (breaking time-reversal symmetry) in the dynamics enables to employ kinetic aspects  that otherwise remain invisible under thermal equilibrium. For instance, recent works have explored how dynamical activity, and more in particular time-symmetric reactivities, can significantly contribute to selection, recovery, or self-assembly.  That includes the more recent references \cite{timematters, non_dissipative_effects, frensteering, kinetic_uncertainty_relation}, but there are of course  the Landauer blowtorch theorem \cite{inadequacy_of_entropy, hb}, or examples such as kinetic proofreading \cite{kinetic_proofreading}, that made the point much earlier.\\ 
The statistical or mean force on a probe coupled to a nonequilibrium medium is likewise influenced by time-symmetric reactivities as first shown in \cite{nongrad}. In the present paper, we continue that study in constructive ways: it is possible to make the probe travel basically any trajectory by correctly choosing the time-symmetric reactivities for every moment during the trajectory. In that fashion, slower dynamical degrees of freedom are steered by an underlying and faster microscopic nonequilibrium dynamics.  More specifically, it will be the phase between entropy flux and frenesy that provides the steering. Such frenetic navigation offers a new tool that works away from equilibrium without the need to adapt energy landscapes.  It is therefore a promising road to be explored, also within the more general context of biological functioning.  The latter is not the subject of the present paper but we sketch the broader context of control and the possible relation with bio-engines in Section \ref{conte}.   For a broader introduction to frenesy we refer to the reviews \cite{fren,nongrad} and to Section \ref{conte} and the Appendix \ref{appendix:cos_phi_dependence} more in particular.\\
%As an example, we can have in mind the recall of memory \cite{frenpaper to be}, but it tells as well about the possible physical origin of rotational forces, limit cycles and chaos as will be the subject of what follows. It also introduces a possible simulation tool where the control variables remain essentially discrete.\\ 

For a more formal introduction, we still add here the main line of strategy and we explain the notion of mean force on a probe (or collective variable). The probe is slow, allowing for a quasistatic approximation where the medium is always in its {\it instantaneous} stationary condition.  We allow the medium to be influenced by the probe's position but not by its velocity:  denoting the probe position by $x$, the medium (only) enters through its stationary probability distribution  $\rho(x,\eta)$ where medium variables are written as $\eta$.  Given a potential $U(x,\eta)$ through which the medium and the probe are coupled, the induced force on the probe is:
\begin{equation}\label{gred}
f(x) =  -\langle \nabla_x U(x,\eta)\rangle(x)= -\sum_\eta\nabla_x U(x,\eta)\rho(x,\eta)
\end{equation}
where $\nabla_x$ is the spatial gradient and the expectation $\langle\cdot\rangle(x)$, like the sum, is over the $\eta$ (assumed discrete) for fixed $x$.\\
 The dynamics of the medium is modeled using a Markov jump process. The rate to jump from $\eta\longrightarrow \eta'$ is
\begin{equation}\label{kk}
k(x, \eta, \eta')=a(x, \{\eta, \eta'\})\exp\big(\frac{\beta}{2}(U(x, \eta) - U(x, \eta') + W(\eta, \eta'))\big)
\end{equation}
where $a(x, \{\eta, \eta'\})=a(x, \{\eta', \eta\})$ is a reactivity or time-symmetric activity parameter, $\beta = (k_BT)^{-1}$ relates to an ambient temperature $T$, and $W(\eta, \eta') = - W(\eta', \eta)$ is the work done by driving forces on the medium when transiting from $\eta$ to $\eta'$.  The fact that $W$ is not written as the difference of a potential indicates the nonconservative nature from which nonequilibrium features emerge.\\
The transition rates \eqref{kk} determine the stationary probability distribution $\rho(x,\eta)$ (assumed unique for now) for making the mean force $f$ in \eqref{gred}.
The main subject of the paper is to understand how to modify that force \eqref{gred} by controlling the $a(x, \{\eta, \eta'\})$ at fixed driving. That means that we do not touch the ``fuel'' (the nonconservative driving giving $W$) or the potential $U$ in the medium; we only manipulate the time-symmetric activities.
The equation of motion of the probe is assumed here to be of the form
\begin{equation}\label{eom}
\dot x= f(x)
\end{equation}
where $x$ lives on some manifold such as the multidimensional torus or the Euclidean space.  We ignore (thermal) noise.  However, the main strategy of controlling the   $a(x, \{\eta, \eta'\})$ does not alter when adding other terms to \eqref{eom}.
In particular, a phase will enter those activity parameters that varies on the time-scale of the slow variables, so that the mean force in \eqref{gred} will use the instantaneous stationary probability. That can be seen as a problem of inversion as is common in control theory, \cite{control_theory,astrom}; we must find the phase shifts in the activity parameters that create the wished-for trajectory in \eqref{eom}.\\

\noindent {\bf Plan of the paper:}  In Section \ref{sec:essence} we recall a strategy to create a rotational component in the mean force.  It is based on \cite{nongrad}, and we indicate how the technique can be  extended to higher dimensions as well.\\  Section \ref{sec:steering_the_force} presents an algorithm to deterministically steer a probe.  It is our main result. Other models or algorithms are conceivable but we make the idea very specific by selecting one of them.\\  In Section \ref{sec:illustrations} we apply that steering technique  for reproducing the trajectories of the van der Pol oscillator and the Lorenz model.\\ In Section \ref{conte} (before the conclusions in Section \ref{sion}) we come back to the general context of the paper, which allows some remarks about the connection with control theory.  We also remind the reader there about the more general notion of frenesy and how it combines with control.  That Section \ref{conte} can also be read before the more technical part starts (in the next sections).\\ The Appendix has four parts  where  we elaborate as well on possible modifications of the nonequilibrium dynamics and on details for the higher-dimensional extensions and rescaling of the force.
\\
For the calculations we use Python \cite{python} with the libraries Numpy\cite{numpy}, Matplotlib \cite{matplotlib} and Scipy \cite{scipy}.

\section{Generating a rotational component in the mean force}
\label{sec:essence}
A first observation about frenetic steering was made in \cite{nongrad} which we repeat here in the same language and for the simplest example.

\subsection{On the unit circle}
\label{sec:rotational_one_d}
Suppose that the probe position is on the unit circle, $x \in S^1$, 
%, and with dynamics
%\begin{equation}\label{ff}
%\dot{x}(t) = f(x(t)) + \xi(t)
%\end{equation}
%where $\xi(t)$ is some mean-zero noise. One should think of \eqref{ff} as an effective dynamics induced by contact with a nonequilibrium medium; see 
undergoing a mean force $f$ defined in \eqref{gred}.   
%for the nature of the force $f$. 
In the present section, we restrict ourselves to the origin of the rotational part $\oint f(x)\id x$, to produce and to indeed modify the rotation of the probe by altering time-symmetric activity parameters of the dynamics of the medium. 

%the probability current of the probe, which is proportional to  \textcolor{blue}{(Is this true?)}

If the medium is in equilibrium for every position $x$, we get for \eqref{gred} a force $f$ which is the gradient of a potential (the free energy). More precisely, when the density $\rho(x, \eta)$ corresponds to canonical equilibrium at inverse temperature $\beta = (k_BT)^{-1}$ for the potential $U(x,\eta)$ for every $x$, we get for the force,  
\begin{align}
f_\text{eq}(x) &= -\frac 1{Z(x)}\sum_\eta\nabla_x U(x,\eta)\exp(-\beta U(x,\eta)), & Z(x)&:=\sum_\eta\exp(-\beta U(x,\eta))\nonumber\\
&= -\nabla_x {\cal F}_\text{eq}(x),& {\cal F}_\text{eq}(x) &:=-k_BT \log Z(x)\nonumber
\end{align}
where $Z(x)$ is the canonical partition function and ${\cal F}_\text{eq}(x)$ is the free energy. Then, the steady behavior is time-reversal symmetric and $\oint f(x)\id x =0$ and there is no current.  It implies that, to have a nonzero rotational component to the force \eqref{gred}, detailed balance has to be broken in the medium, but even that is not enough, as we shall now see.\\

\begin{figure}
    \centering
    \includegraphics[height=0.3\textheight]{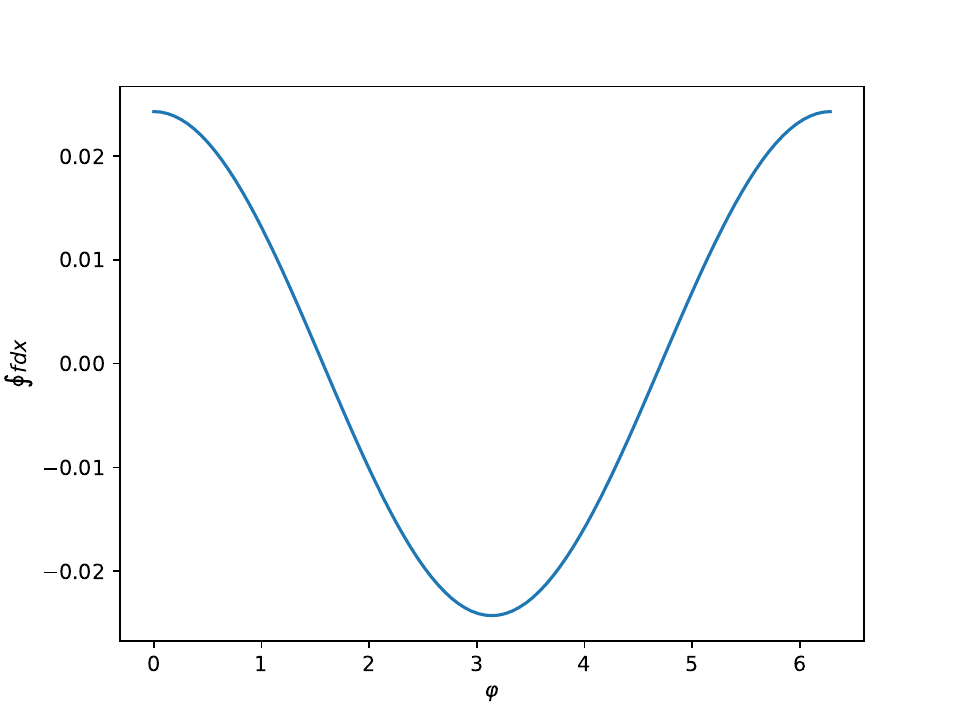}
    \caption{The rotational part of the force for the system described around \eqref{ex1} with $\beta=\varepsilon=b=1$ and $\lambda=0.2$: it turns out that  $\oint f\,\id x \approx 0.0243\,\cos \varphi$, plotted as a function of the shift $\varphi$.  Note in particular that $\oint f\id x = 0$ when $\varphi = \pm \pi/2$, {\it i.e.}, when the reactivities \eqref{ex1} are a function of the interaction energy $U(x,\eta)$.}
    \label{fig:f_rot}
\end{figure}

To be specific about \eqref{kk}, we take a medium with states $\eta = 1, 2, 3$, and a driving $W(x,1,2)=W(x,2,3) =W(x,3,1)=\varepsilon$ along the loop. The interaction potential is $U(x,1)=\lambda\sin x$ and $U(x,\eta=2,3)=0$. For the reactivities in \eqref{kk} we take:
\begin{align}\label{ex1}
a(x, \{1,2\})&=1+\lambda b\cos(x+\varphi),\qquad a(x, \{1,3\})=1-\lambda b\cos(x+\varphi)\\
a(x, \{2,3\})&=1\nonumber
\end{align}
which again contain the coupling parameter $\lambda > 0$, and we take an additional $b>0$ for changing the coupling in the reactivities.
The force \eqref{gred} becomes 
\begin{equation}
f(x) = -\lambda\,\rho(x,1)\,\cos x 
\end{equation}
The idea is to trap the system in the state $\eta$ that provides the required force. By making $a(x,\{1,2\})$ small, we trap the medium in state 1, which means to increase $\rho(x,1)$. For example with $\varphi=0$, $a(x, \{1, 2\})$ is at its minimum for  $\cos x =-1$.  For those values of $x$ the $-\lambda\,\cos x$ factor in the force is maximal, so that we get a positive value for $\oint f(x)\id x$. Manipulating the shift $\varphi$ as a control, we can change for which values of $x$ the reactivity $a(x, \{1, 2\})$ is small and so influence $\oint f(x)\id x$.\\  
%Note that not only the potential but also the activity parameters depend on $x$ \textcolor{blue}{(Find out if this is essential. We know in weak coupling it is but is it also like that in general?)}. 
In Fig.~\ref{fig:f_rot} the dependence of $\oint f(x)\id x$ on $\varphi$ is shown for $\beta=\varepsilon=b=1$ and $\lambda=0.2$; we see that $\oint f(x)\id x\, \simeq 0.0243\cos \varphi$. As shown in greater generality in Appendix \ref{appendix:cos_phi_dependence}, to second order in $\lambda$, the rotational part of the force is indeed proportional to $\cos \varphi$.  The important conclusion here is that the rotational part of the force $\oint f(x)\id x$ can be controlled with $\varphi$. Note that the above model can of course be run on the line $\mathbb{R}$, giving a periodic force $f(x)$, and instead of rotation, a drift to infinity.
In accordance with the major conclusion in \cite{nongrad}, and as obvious by inspecting Fig.~\ref{fig:f_rot}, the $x$-dependence of the interaction energy and the $x$-dependence of the activity parameters must have a relevant phase difference.  That shift will be the major tool for steering in what follows.\\

We extend the above construction in the next section to multiple dimensions (and higher-dimensional tori).  Note however that the rotational part $\oint f(x)\,\id x$ does not completely determine the motion.  That gets remedied in Section \ref{sec:steering_the_force}.

\subsection{Multiple dimensions}\label{md}
\begin{figure}
    \centering
    \includegraphics[height=0.3\textheight]{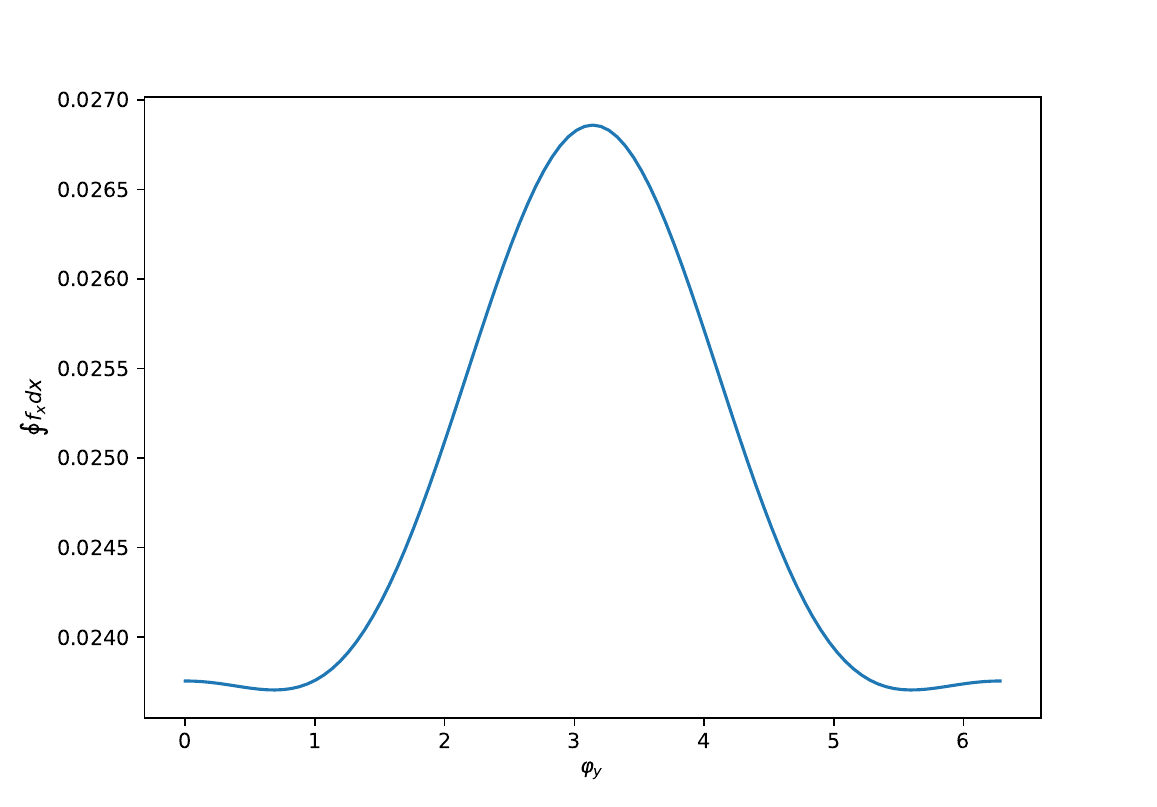}
    \caption{The system described around \eqref{aha} at $\lambda=0.2$, $\beta=\varepsilon=b_x=b_y=1$ and $\varphi_x=0$: $\oint f_x\id x > 0$ evaluated at $y=0$ is almost constant as a function of $\varphi_y$.  The variation is of order less than $\lambda^2$.}
    \label{fig:f_rot_x_phi_y}
\end{figure}

We start from the same setup as above in \eqref{ex1} but we add equivalent terms for a second dimension with variable $y\in S^1$.  We still have a medium with states $\eta = 1, 2, 3$ and a driving $W(1,2)=W(2,3) =W(3,1)=\varepsilon$, independent of $(x,y)$. For the energy we now take $U(x, y, 1)=\lambda\sin x + \lambda\sin y$ and $U(x,y,\eta=2,3)=0$.  For the activity parameters we put
\begin{align}\label{aha}
    a(x, y, \{1,2\})&=1+\lambda b_x\cos(x+\varphi_x)+\lambda b_y\cos(y+\varphi_y)\\
    a(x, y, \{1,3\})&=1-\lambda b_x\cos(x+\varphi_x)-\lambda b_y\cos(y+\varphi_y)\nonumber\\
    a(x, y, \{2,3\})&=1\nonumber
\end{align}
The force \eqref{gred} then has two components 
\begin{eqnarray*}
f_x(x,y) &=& -\lambda\,\rho(x,y,1)\,\cos x\\
f_y(x,y)&=& -\lambda\,\rho(x,y,1)\,\cos y
\end{eqnarray*}
Starting from results in \cite{nongrad}, we demonstrate in Appendix \ref{appendix:multiple_dimensions} that to second order in $\lambda$, $\oint f_x(x, y)\id x$ is not influenced by the $y-$dependencies in $U$ or in the $a(x,y)$'s (including $b_y$),  
%$b_y$ or $\varphi_y$ \textcolor{olive-green}{(Also not the $\lambda\sin y$ part of the potential)}
and similarly for $\oint f_y(x, y)\id y$. An example of that is shown in Fig.~\ref{fig:f_rot_x_phi_y} where we indeed see that the influence of $\varphi_y$ on $\oint f_x(x, y)\id x$ is limited. That means that we can control $\oint f_x(x, y)\id x$ using $\varphi_x$ just like in the one-dimensional case without worrying that the $y$-dependency in $U$ and in the $a(x,y)$'s would prevent us from choosing the sign of $\oint f_x(x, y)\id x$.  That obviously makes the steering much easier.
%Of course the particle's $y$ position would change when considering the time evolution of the particle but we can alter $\oint f_x(x, y)\id x$ including changing its sign by altering $\varphi_x$ (only).

\section{Steering the force}\label{sec:steering_the_force}

%Our task here is to produce a prescribed trajectory $\dot{x}_t$ by given a time-dependent protocol in the activity parameters. Of course, $\dot{x}_t = f(x_t)$ and therefore, we want to create a given force for every position on the circle. Here we describe such a model starting again with a probe moving on the unit circle, $x_t\in S^1$.\\

Turning the direction of the front wheel of a bicycle is quite useless when standing.  Once in motion however, steering the bicycle becomes possible by controlling the handlebar or leaning the bike.  In the following, we similarly manipulate time-symmetric aspects to create the required force.
%Directing the motion of cattle or sheep is perhaps an even older task than irrigation. There is evidence of the use of herding dogs since about 6,000 years ago.  Clearly, those {\it active} dogs do not use any gradient in slope but they impose the required motion by coupling the sheep position to their own frenziness. 

\subsection{One dimension}
\label{sec:steering_the_force_one_d}
We start again with the situation of a probe moving on the unit circle as the result of a force induced by the interaction of the probe with a nonequilibrium medium. In the previous section, we could only control $\oint f(x)\id x$. Now, remaining in the quasistatic approximation, ignoring fluctuations, we want to  control the complete force $f(x)$ in \eqref{gred}, determining the dynamics of the probe,
\begin{equation}
\dot{x}(t)=f(x(t))
\end{equation}
We choose here for an overdamped description, keeping the interpretation that $x$ refers to a position and $f$ is a force.  That is typical for a biological context, but obviously, it can also refer to a much broader setup as used in the theory of dynamical systems (see Section \ref{sec:illustrations}).\\

It actually suffices to take 3 states in a cycle for the medium, $\eta\in\{0, 1, 2\}$.  We refer to \eqref{kk} for the structure of transition rates $k(x, \eta,\eta')$. There is a driving with constant magnitude $W(0, 1)=W(1, 2)=W(2, 0)=\varepsilon$. 
We use the interaction energy $U(x, \eta)=\lambda\sin(x-\eta2\pi/3)$ and for activity parameters we put
\begin{equation}\label{act3}
a(x, \{\eta, \eta'\})=\left\{
\begin{array}{rl}
1+\lambda b\cos(x-\eta2\pi/3+\varphi)&\text{if } \eta'-\eta\equiv 1\pmod{3}\\
1+\lambda b\cos(x-(\eta-1)2\pi/3+\varphi)&\text{if } \eta'-\eta\equiv 2\pmod{3}
\end{array}\right.
\end{equation}
The ``shift'' $\varphi$ will control $f(x)$ for every $x$.\\

\begin{figure}
    \centering
    \includegraphics[height=0.3\textheight]{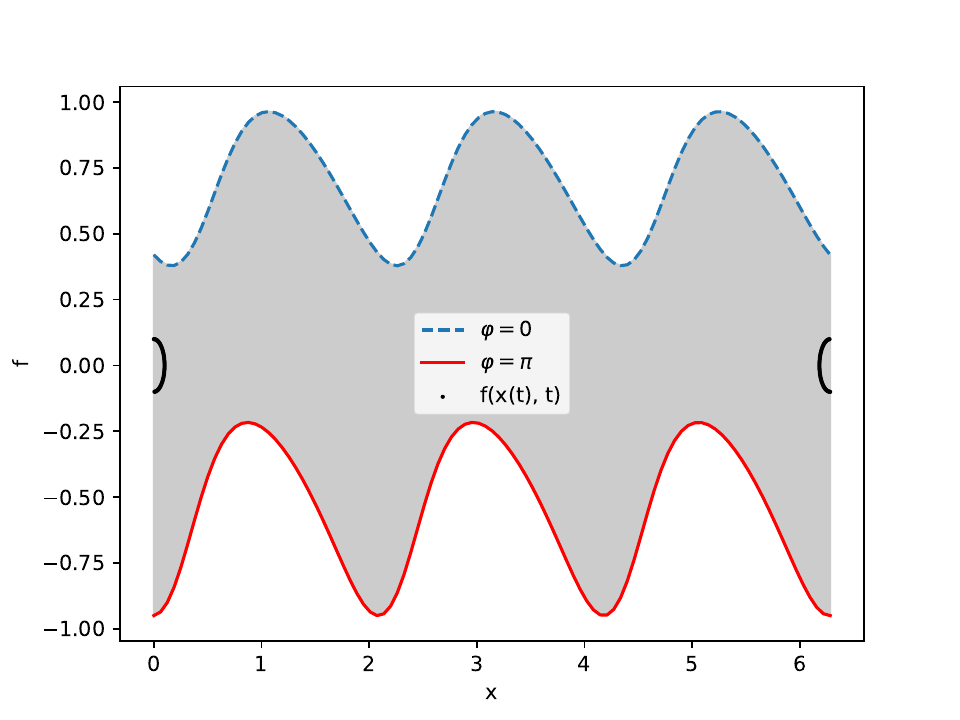}
    \caption{$f(x)$ for $b=0.99,\ \lambda=1,\ \beta=1\ \text{and }\varepsilon=5$, for $\varphi=0$ and $\varphi=\pi$. The shaded region indicates that any value of the force in that region is attainable by a suitable choice for $\varphi$. The black dots are values of the (time-dependent) force $f(x(t), t)$ for the trajectory $x(t)=0.1\sin(t)$ for 100 data points between $t=0$ and $t=2\pi$.}
    \label{fig:f_2_phis_3_states}
\end{figure}

As we see in Fig.~\ref{fig:f_2_phis_3_states}, when $\varphi=0$ the force is positive everywhere. As for the model in Section \ref{sec:essence}, that is achieved by trapping the system in the state that provides a positive force, but now the system has to be trapped in different states depending on the position. Using more than 3 states is possible. In case we use more than 3 states (when taking the same period for the sine and cosine functions) the curves in Fig.~\ref{fig:f_2_phis_3_states} would have a smaller amplitude and a smaller period. In the rest of the text we use 3 states for the model of the medium. In Appendix~\ref{app:4_states} we discuss results for 4 states.\\
Fig.~\ref{fig:f_2_phis_3_states} shows the force on the probe for $\varphi=0\text{ and }\varphi=\pi$ depending on its position $x$ on the circle, and for parameter values $b=0.99,\ \lambda=1,\ \beta=1\ \text{and }\varepsilon=5$. Depending on the shift $\varphi$, we can thus obtain either strictly positive or strictly negative values for that force. As it depends on $\varphi$ in a continuous way, for every position on the circle we can also attain every value for the force between that positive and negative value. 
%in any desired symmetric interval.  
That gives a way to steer deterministically simply by controlling $\varphi$ at preset values of (large enough) $\varepsilon$.\\ 
%Note that the assumption that the probe moves on the unit circle ($x_t\in S^1$) is not essential and the model works just as well on the line $\mathbb{R}$. We simply initially created the model with rotation in mind. More concretely we can think of two applications. 

\begin{figure}
    \centering
    \includegraphics[height=0.3\textheight]{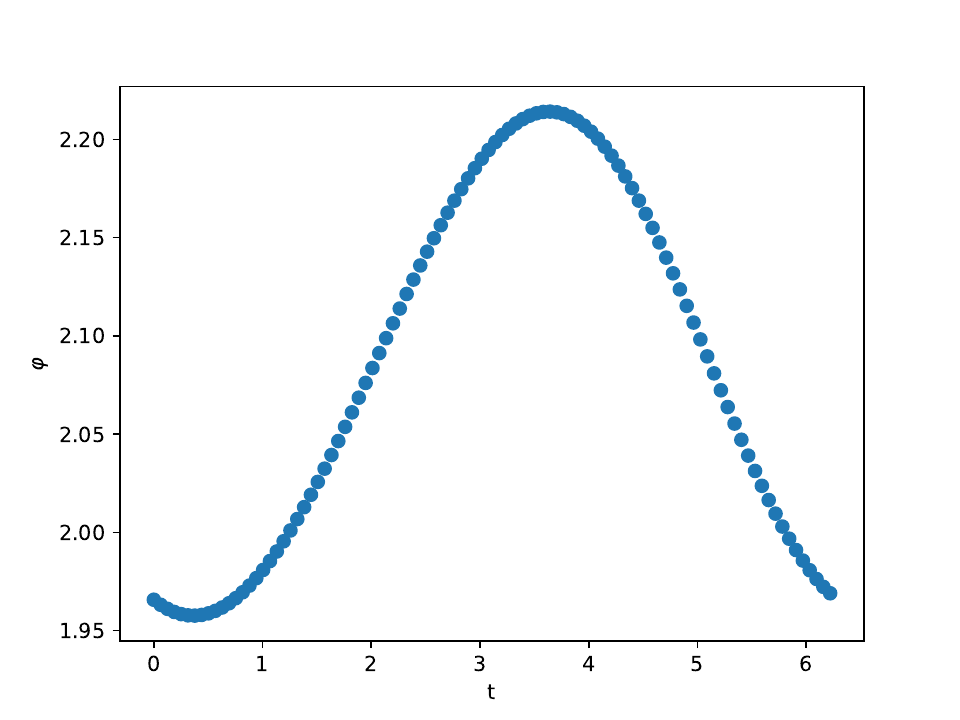}
    \caption{$\varphi(t)$ for the trajectory $x(t)=0.1\sin(t)$ for 100 data points between $t=0$ and $t=2\pi$ and for parameter values $b=0.99,\ \lambda=1,\ \beta=1\ \text{and }\varepsilon=5$.}
    \label{fig:phi_t_3_states}
\end{figure}

Given a trajectory $x(t), t\in [0, T]$ for which $\dot{x}(t)$ is between the two curves in Fig.~\ref{fig:f_2_phis_3_states}, we can find $\varphi(t)$ so that the trajectory $x(t)$ is the result of the force on the probe induced by the medium: $\dot{x}(t)=f(x(t), \varphi(t))$.\\ 
As illustration, Fig.~\ref{fig:phi_t_3_states} shows $\varphi(t)$ for the given trajectory $x(t)=0.1\sin t$ for 100 data points between $t=0$ and $t=2\pi$. Parameter values $b=0.99,\ \lambda=1,\ \beta=1\ \text{and }\varepsilon=5$ were used again. $f(x(t), t)$ for that trajectory $x(t)$ is shown in Fig.~\ref{fig:f_2_phis_3_states}, also for 100 data points between $t=0$ and $t=2\pi$. 

%\textcolor{olive-green}{(Is dit in detail genoeg of moet ik hier een uitleg doen met een voorbeeld van een derdemachtsveelterm en zijn snijpunten met een constante functie? Het is ook zo dat het algoritme niet de oplossing geeft die het dichtst ligt bij de vorige oplossing, maar dat blijkt in vele gevallen wel zo te zijn.)}\\ 
%It seems that for the algorithm we use there is a jump in $\varphi$ between around 1.6 to 1.9. 
If $\dot{x}(t)$ has a broader range of values (not between the two curves in Fig.~\ref{fig:f_2_phis_3_states}), then we can apply a rescaling: we write $x=Ax'$ for some $A>1$, big enough, so that $\dot{x}'(t)$ lies again between the two curves in Fig.~\ref{fig:f_2_phis_3_states}, supposing that $\dot{x}(t)$ is finite. We can then of course choose $\varphi(t)$ for the trajectory $x'(t)$ and retrieve $x(t)$ from $x'(t)$: $x(t)=Ax'(t)$. Note that $\varphi(t)$ depends on the rescaling and that that dependence is not trivial. 
%A good choice for $a$ would be $\max \big\{|\dot{x}(t)|\,\big|t\in [0, T]\big\}/0.2$ (notice in Fig.~\ref{fig:f_2_phis} that for the curve for $\varphi=0$: $f(x)>0.2$ and that for the curve for $\varphi=\pi$: $f(x)<-0.2$). Note that the values for $\varphi(t)$ depend on the choice for the scale $a$. 
%We explain the details in Appendix \ref{sca}.

A second application is that given a dynamical system with equation $\dot{x}=g(x)$, where $g(x)$ lies between the two curves in Fig.~\ref{fig:f_2_phis_3_states}, we can find $\varphi(x)$ so that $f(x, \varphi(x))=g(x)$. If $g(x)$ has a broader range of values, in many cases we can find an interval, say $[x_0, x_1]$, in which $g(x)$ is bounded. We can then apply a rescaling $x=Ax'$ in a similar way as above. We explain the details in Appendix \ref{appendix:scaling}.

\subsection{Higher-dimensional steering}
In order to steer deterministically in multiple dimensions we cannot just add the perturbations for each dimension as we did in Section \ref{md}.  
%something similar to what we did in the previous section. In this model we try to trap the system in the state that delivers the desired force. If we would do something similar to what we did in the previous section, then once for a certain position $(x,y)$, we have $\rho(x,y,\eta)$ so that we have the desired $f_x(x,y)$, then $f_y(x,y)$ would be determined as well and so would not be able to be chosen by us. 
The appropriate extension to higher dimensions is to view the three medium states (of Section \ref{sec:steering_the_force_one_d}) as the values of a spin and to have a spin for every dimension. One could say that every spin is responsible for the force along its dimension and the spins are all independent (noninteracting).  The interaction energy with the probe is additive.\\
Let us denote the state of the spin for dimension $x$ as $\eta^x$.
We take 2 dimensions for notational simplicity. 
We already know that the following should hold:  $\rho(x, y, \eta^x, \eta^y)=\rho(x, \eta^x)\rho(y,\eta^y)$ and 
\begin{align}
f_x(x,y)&=-\langle\partial_xU(x, y, \eta^x, \eta^y)\rangle(x,y)\\
&=-\langle\partial_xU_x(x, \eta^x)\rangle(x)\nonumber\\
&=-\sum_{\eta^x}\partial_xU_x(x, \eta^x)\rho(x, \eta^x)\nonumber
\end{align}
but we recall the dynamics:\\

For 2 dimensions, we have two spins $\eta^x, \eta^y \in \{0, 1, 2\}$. We consider separate Markov processes for every dimension, and we use the transition rates as specified under \eqref{kk} and around \eqref{act3}.\\
The two spins have the same driving (as before): e.g., $W_x(0, 1)=W_x(1, 2)=W_x(2, 0)=\varepsilon$ and  the same for the driving on $\eta^y$. Both spins have an energy: $U_x(x, \eta^x)=\lambda\sin(x-\eta^x2\pi/3)$ and similarly for $\eta^y$ and the interaction energy with the probe is the sum $U(x,y,\eta^x,\eta^y)=U_x(x, \eta^x)+U_y(y, \eta^y)$. For the activity parameters, we have for the jump between values $\eta^x_i$ and  $\eta^x_j$
\begin{equation}\label{aii}
a_x(x, \{\eta^x_i, \eta^x_j\})=\left\{
\begin{array}{rl}
1+\lambda b\cos(x-\eta^x_i2\pi/3+\varphi_x)&\text{if } \eta^x_j-\eta^x_i\equiv 1\pmod{3}\\
1+\lambda b\cos(x-(\eta^x_i-1)2\pi/3+\varphi_x)&\text{if } \eta^x_j-\eta^x_i\equiv 2\pmod{3}
\end{array}\right.
\end{equation}
and again similarly for $\eta^y$. 

Given a trajectory $(x(t), y(t)), t\in[0, T]$, we  find $\varphi_x(t)$ and $\varphi_y(t)$ independently and in the same way as in  one dimension, by requiring $\dot{x}(t)=f_x(x(t), \varphi_x(t))$ and $\dot{y}(t)=f_y(y(t), \varphi_y(t))$. In other words, given a required force at a certain position $(x,y)$ with components $g_x(x,y)$ and $g_y(x,y)$, we find $\varphi_x(x,y)$ and $\varphi_y(x,y)$ independently in the same way as in  one dimension with $f_x(x, \varphi_x(x,y))=g_x(x,y)$ and $f_y(y, \varphi_y(x,y))=g_y(x,y)$.\\ 
There remains the issue that rescaling may be needed depending on the range of the force.

% That means, given a dynamical system with the following equations:
% \begin{align}
% \dot{x}&=g_x(x,y)\nonumber\\
% \dot{y}&=g_y(x,y)\nonumber
% \end{align}
% , we can try to find $\varphi_x(x,y)$ and $\varphi_y(x,y)$ so that $f_x(x,y)=g_x(x,y)$ and $f_y(x,y)=g_y(x,y)$. We still have the issue that the force we can generate with our model is bounded, but, if we can find an area in which $g_x$ and $g_y$ are bounded, we can do a rescaling of the system in a similar way as before so that for that area we succeed in finding $\varphi_x(x,y)$ and $\varphi_y(x,y)$. This rescaling is explained in more detail in Appendix \ref{sca}. 

\section{Illustrations}
\label{sec:illustrations}
%Forces derived as gradient of a potential have a limited phenomenology......
When the medium is in equilibrium and there is a fixed interaction potential $U(x, \eta)$, then the force is fixed and conservative: $f(x)=-\nabla_x\mathcal{F}_\text{eq}(x)$. This seriously limits the possibilities for systems where the force is modeled as the force induced by the interaction of a probe with an equilibrium medium. When the medium is nonequilibrium the force can be nonconservative and additionally it is not completely determined  by the interaction potential $U(x, \eta)$; we can modify the time-symmetric activity parameters of the dynamics of the medium to depend on the position $x$ and time.\\

In Section \ref{sec:steering_the_force} we have introduced a model for the medium where we altered the time-symmetric activity parameters by means of  the shift $\varphi$; see \eqref{aii}. In what follows we use that algorithm to reproduce the dynamics of the van der Pol oscillator and the Lorenz system. That is achieved by using the 3-state cycle model described in Section \ref{sec:steering_the_force}, where we always employ parameter values $b=0.99,\ \lambda=1,\ \beta=1\ \text{and }\varepsilon=5$.\\
Say for two dimensions like for the van der Pol oscillator, we start from given equations of motion,
\begin{align}
\dot{x}&=g_x(x,y)\label{eq:dynamical_system_2d}\\
\dot{y}&=g_y(x,y)\nonumber
\end{align}
For the probe to follow that evolution, we use two approaches.\\
In the first approach, we select a region of state space on which we put a discrete grid.  For all points $(x_g, y_g)$ in the grid, we calculate $\varphi_x$ so that $f_x(x, \varphi_x(x_g,y_g))=g_x(x_g,y_g)$ and similarly for $\varphi_y$. Then, given $(x, y)$, we find the nearest grid point $(x_g, y_g)$  and get $\varphi_x(x_g, y_g),\varphi_y(x_g, y_g)$, after which we calculate $f_x(x_g, \varphi_x(x_g, y_g))$ and $f_y(y, \varphi_y(x_g, y_g))$, which would be good approximations to  $g_x(x,y)$ and $g_y(x,y)$. \\
In the second approach we numerically integrate the equations \eqref{eq:dynamical_system_2d} for $t\in [0, T]$ given $x(0)$ and $y(0)$ and for each time step we calculate $\varphi_x(t)$ so that $f_x(x(t), \varphi_x(t))=g_x(x(t), y(t))$ and similarly for $\varphi_y(t)$. That way our nonequilibrium model produces approximately the same trajectory for the same $x(0)$ and $y(0)$ as the dynamical system. \\
For both these approaches it is possible to use a parameterized functional form for $\varphi_x$ and $\varphi_y$ so that a fitting procedure would drastically decrease the storage of data. \\
Generalizing to an arbitrary number of dimensions is straightforward. 

\subsection{van der Pol oscillator}
\begin{figure}
    \centering
    \includegraphics[height=0.3\textheight]{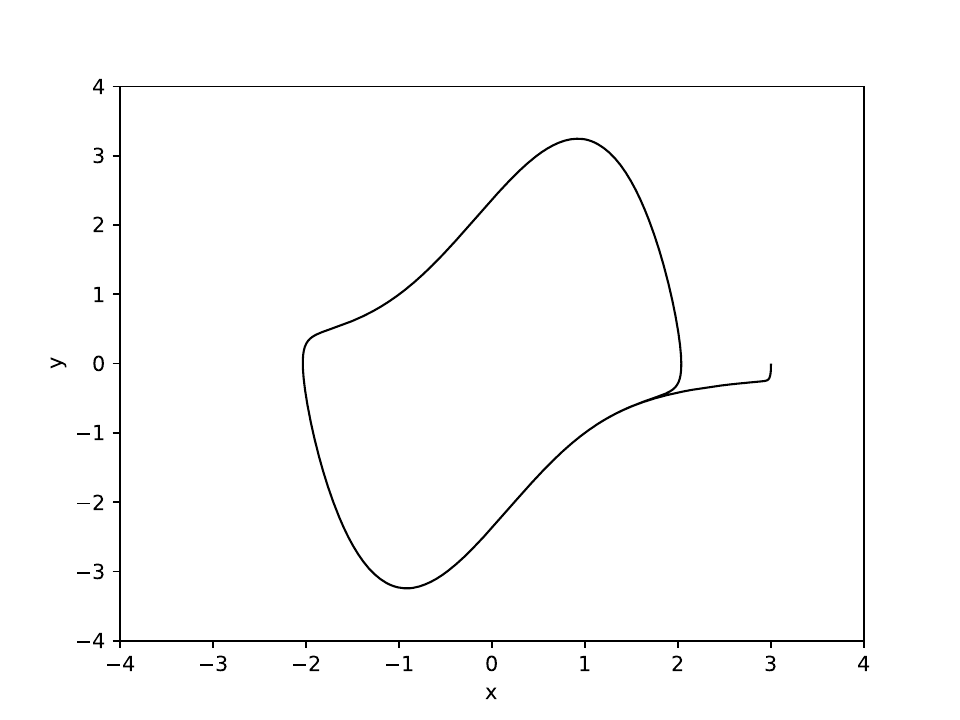}
    \caption{A trajectory for a probe that approximates the van der Pol oscillator, for damping $\mu=1.5$, initial state $x=3, y=0$, time step $0.01$ and duration $T=12$.}
    \label{fig:van_der_pol_steering_3_states}
\end{figure}

\begin{figure}
    \centering
    \includegraphics[height=0.3\textheight]{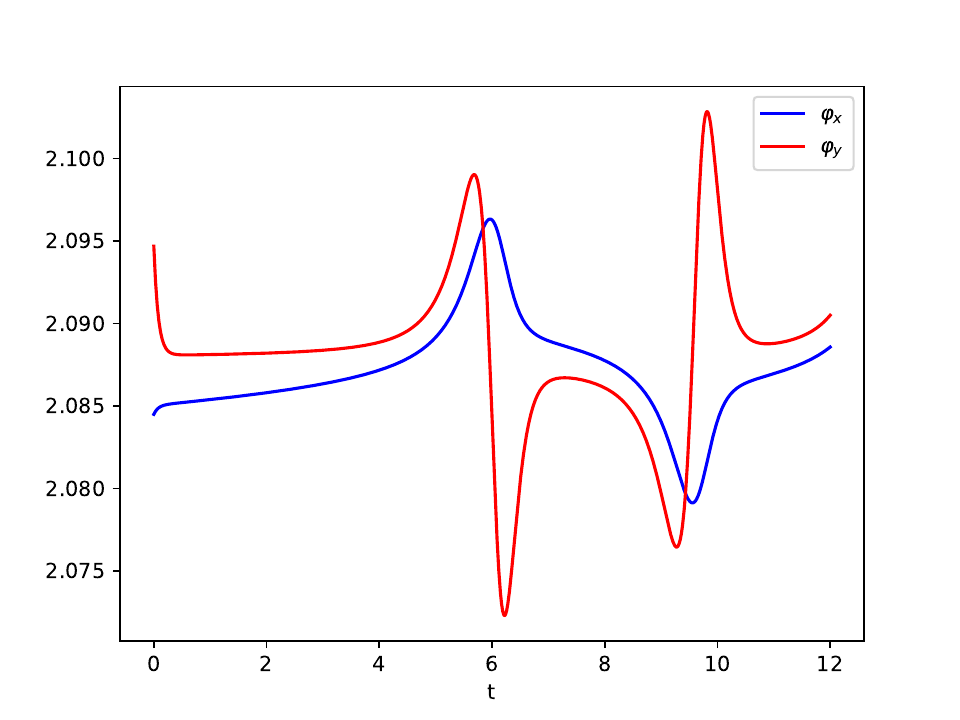}
    \caption{Phase shifts $\varphi_x(t)$ and $\varphi_y(t)$ for the medium dynamics coupled to the probe that follows the van der Pol oscillator, for $\mu=1.5$, initial state $x=3, y=0$, and time step $0.01$.}
    \label{fig:van_der_pol_phi_t_3_states}
\end{figure}

The equations for the van der Pol oscillator \cite{strogatz, vanderpol} are
\begin{align}
\dot{x}&=y\\
\dot{y}&=-\mu(x^2-1)y-x\notag
\end{align}
modeling oscillations with nonlinear damping and amplification.  It was originally (and is still) used in the context of electric activity, in particular in biological systems \cite{vanderpolmark}.\\
%\cite{van der Pol and van der Mark (1928)}.\\
We take $\mu=1.5$, and the region $[-4, 4]\times[-4, 4]$, for which the rescaling $A=500$ is appropriate.  We use 1000 grid points in both dimensions giving a total of $10^6$ grid points. Using the dynamics with transition rates described in \eqref{kk}, \eqref{act3} and \eqref{aii}, we plot a trajectory for initial state $x=3, y=0$, time step $0.01$ and duration $T=12$, shown in Fig.~\ref{fig:van_der_pol_steering_3_states}. That is exactly what we expect that trajectory to look like for the van der Pol oscillator. The shifts $\varphi_x(t)$ and $\varphi_y(t)$ calculated for the same values for the parameters (so for initial state $x=3, y=0$, time step $0.01$ and $T=12$) are shown in Fig.~\ref{fig:van_der_pol_phi_t_3_states}.\\

If $\mu=0$, the solution would be $x(t)=3\cos(t)$ and so a translation and rescaling of the trajectory $x(t)$ for which $\varphi(t)$ is shown in Fig.~\ref{fig:phi_t_3_states} (rescaling not only because of the amplitude of $x(t)$ but also because of $A=500$). %We would expect that in case $\mu=0$ the forces are in a similar range as in Fig.~\ref{fig:van_der_pol_phi_t}. 
Because $x(t)$ for which $\varphi(t)$ is shown in Fig.~\ref{fig:phi_t_3_states} is on a different scale as $x(t)=3\cos(t)$ (with rescaling $A=500$) there is however a big difference in the range of $\varphi(t)$.

% $\varphi_x(t)$ and $\varphi_y(t)$ in Fig.~\ref{fig:van_der_pol_phi_t} should be in the same range as $\varphi(t)$ in Fig.~\ref{fig:phi_t}. In contrast to $\varphi(t)$ in Fig.~\ref{fig:phi_t} there are no jumps here

\subsection{Lorenz model}
\begin{figure}
    \centering
    \subfigure[\,exact dynamics]{
        \includegraphics[height=0.25\textheight]{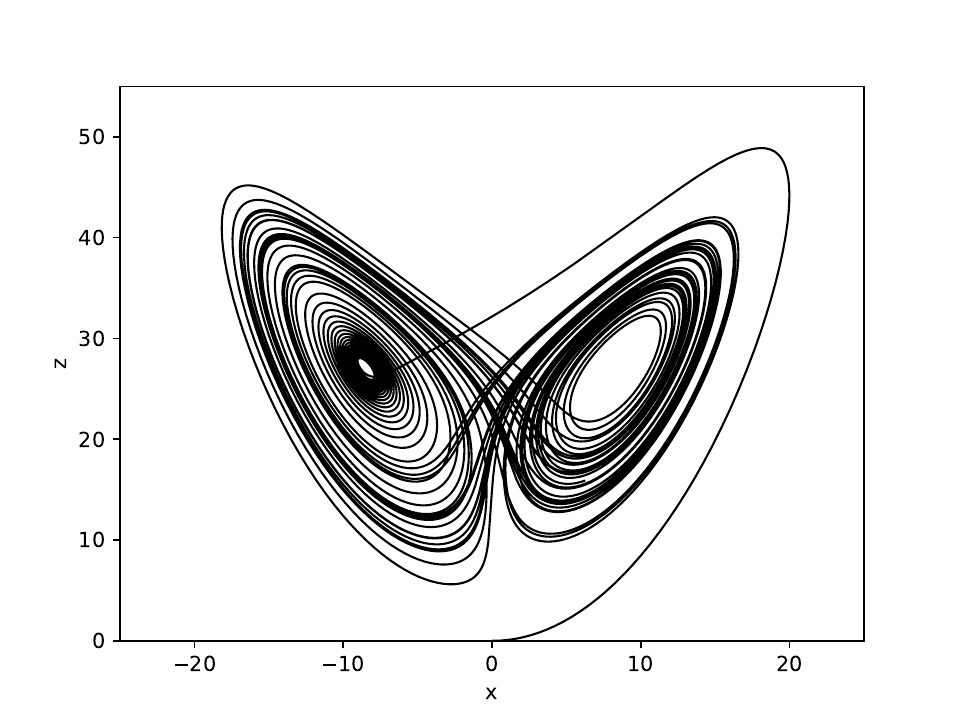}
        \label{fig:lorenz_exact_3_states}
    }
    \subfigure[\,probe dynamics]{
        \includegraphics[height=0.25\textheight]{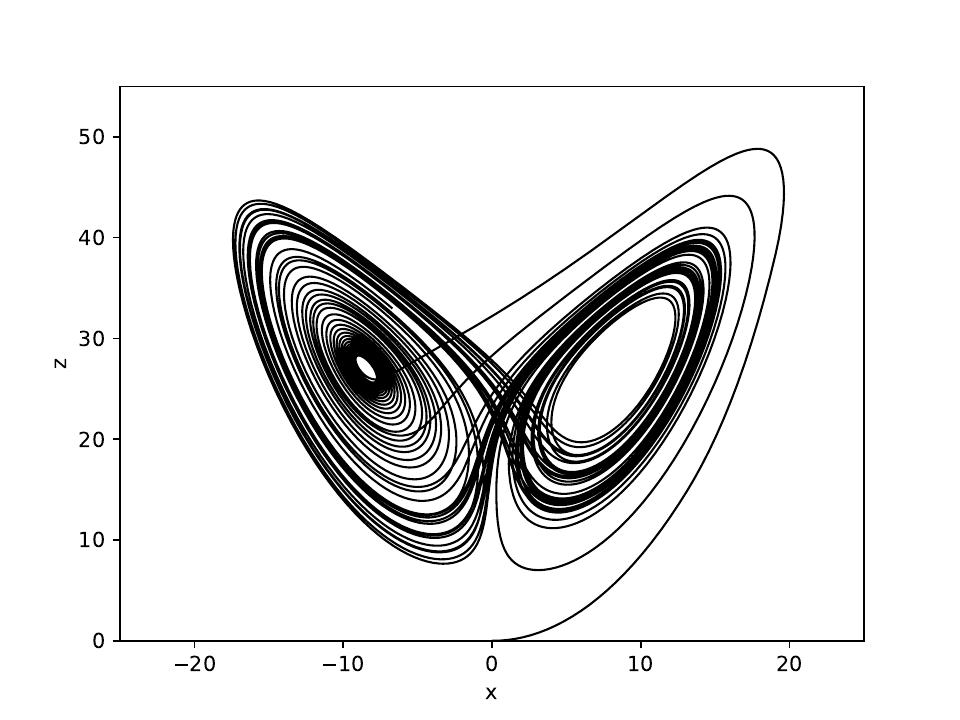}
        \label{fig:lorenz_steering_3_states}
    }
    \caption{Trajectories $(x(t), z(t))$ using the exact dynamics (in (a)) and the approximated dynamics given by our nonequilibrium model (in (b)), for initial state $x=0,\ y=1,\ z=0$, time step $0.001$ and duration $T=50$. }
	\label{fig:lorenz_trajectory_3_states}
\end{figure}

\begin{figure}
    \centering
    \subfigure[\,$\varphi_x(t)$]{
        \includegraphics[height=0.25\textheight]{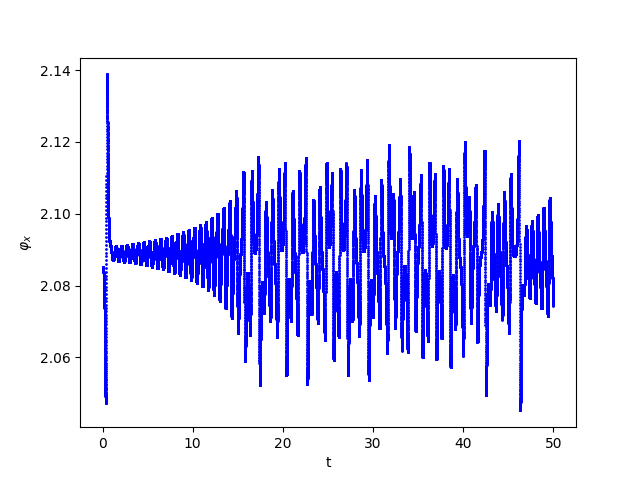}
        \label{fig:lorenz_phi_x_t_3_states}
    }
    \subfigure[\,$\varphi_z(t)$]{
        \includegraphics[height=0.25\textheight]{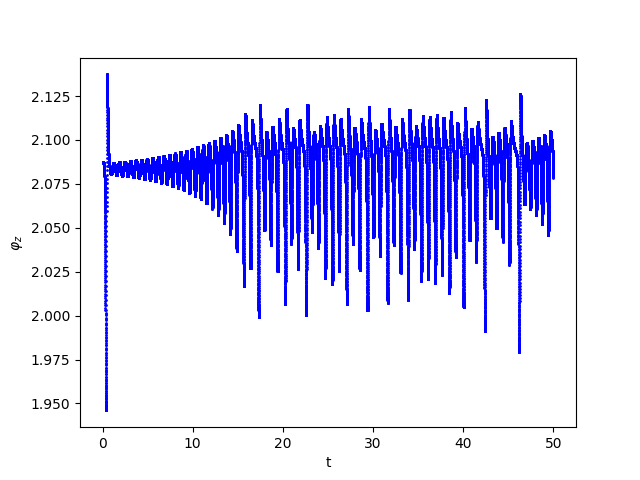}
        \label{fig:lorenz_phi_z_t_3_states}
    }
    \caption{$\varphi_x(t) (a)\text{ and } \varphi_z(t) (b)$ for a dynamics based on our nonequilibrium model that approximates the Lorenz model, for initial state $x=0, y=1, z=0$, time step $0.001$ and $T=50$.}
    \label{fig:lorenz_phi_t_3_states}
\end{figure}

The 60-year-old Lorenz system \cite{strogatz, lorenz} is particularly relevant for our purposes because it is (already) a reduced version of a larger system, \cite{saltzman}.
% \cite{Saltzman, Barry (1962). "Finite Amplitude Free Convection as an Initial Value Problem—I". Journal of the Atmospheric Sciences. 19 (4): 329–341. Bibcode:1962JAtS...19..329S. doi:10.1175/1520-0469(1962)019<0329:FAFCAA>2.0.CO;2.}.  
The derivation, called Oberbeck–Boussinesq approximation, starts from the equations of irreversible thermo- and hydrodynamics, in particular incorporating the phenomenon of  Rayleigh–B\'enard convection.  In the end, using a spectral Galerkin approximation, a set of three coupled, nonlinear ordinary differential equations are obtained, called the Lorenz equations for the real variables $(x,y,z)$,  
\begin{align}
\dot{x}&=\sigma(y-x)\label{eq:lorenz}\\
\dot{y}&=rx-y-xz\nonumber\\
\dot{z}&=xy-bz\nonumber
\end{align}
% \cite{Hilborn, Robert C. (2000). Chaos and Nonlinear Dynamics: An Introduction for Scientists and Engineers (second ed.). Oxford University Press. ISBN 978-0-19-850723-9.}
% \cite{Bergé, Pierre; Pomeau, Yves; Vidal, Christian (1984). Order within Chaos: Towards a Deterministic Approach to Turbulence. New York: John Wiley & Sons. ISBN 978-0-471-84967-4.}
%We use the same parameters and initial conditions as in \cite{strogatz}. 
% Hilborn (2000), Appendix C; Bergé, Pomeau & Vidal (1984), Appendix D; or Shen (2016),[28] Supplementary Materials
%(Shen referentie:) Shen, B.-W. (2015-12-21). "Nonlinear feedback in a six-dimensional Lorenz model: impact of an additional heating term". Nonlinear Processes in Geophysics. 22 (6): 749–764. Bibcode:2015NPGeo..22..749S. doi:10.5194/npg-22-749-2015. ISSN 1607-7946.
The physics derivation is included in texts such as \cite{hilborn, berge, shen}.  Our frenetic steering is obviously also using a reduced description, a much simpler but artificial one.  We ignore to what extent the parameters $\sigma, r$ or $b$ pick up frenetic features in the atmospheric dynamics. \\
We take $\sigma=10,\ r=28,\ b=8/3$ and the region $[-25, 25]\times[-25, 25]\times[0, 55]$, for which the rescaling $A=3700$ is appropriate.  We use 1000 grid points in each direction, making $10^9$ grid points in total. In the case of the van der Pol oscillator, which has two dimensions, it was possible to store all the values for $\varphi_x(x,y)$ and $\varphi_y(x,y)$ for all grid points in memory. In the case of the Lorenz model, which has three dimensions, all this data would have a size of around 80 GB, which exceeds the memory of most PC's. In order to deal with that problem, we determine the nearest grid point and its values for $\dot{x},\ \dot{y},\ \dot{z}$ as given by \eqref{eq:lorenz}.\\
Using the dynamics provided by our nonequilibrium medium (around \eqref{act3}, as extension to three dimensions of \eqref{aii}), a trajectory $x(t), z(t)$ for initial state $x=0,\ y=1,\ z=0$, time step $0.001$ and $T=50$ is plotted in Fig.~\ref{fig:lorenz_steering_3_states}, next to Fig.~\ref{fig:lorenz_exact_3_states} where we see the trajectory for the same parameters but calculated from \eqref{eq:lorenz},  exhibiting the typical Lorenz butterfly for its strange attractor, \cite{warwick}.
% \cite{Tucker Warwick (2002). A Rigorous ODE Solver and Smale's 14th Problem. Foundations of Computational Mathematics. 2 (1): 53–117. CiteSeerX 10.1.1.545.3996. doi:10.1007/s002080010018. S2CID 353254.}.  
Despite some local differences, not very surprising for a dynamics that exhibits chaos, the global features are perfectly reproduced.\\
In Fig.~\ref{fig:lorenz_phi_t_3_states} the shifts $\varphi_x(t)\text{ and }\varphi_z(t)$ are shown for the same values for the parameters (initial state $x=0,\ y=1,\ z=0$, time step $0.001$ and $T=50$). The (irregular) oscillations reproduce the chaotic nature of the trajectories $x(t)$ and $z(t)$.

\section{Frenetic steering and control theory}\label{conte}
The present section collects some further thoughts on frenetic control and broadens the context for a better understanding of the specific nature of the applied schemes.\\

Steering is an extended control, which wants to produce (as time evolves) the required trajectory, not just a final condition.  Nevertheless, the present paper can be situated under the general heading of control theory.  The latter searches for algorithms to engineer the required inputs that drive the system from some initial state(s) to a desired state or condition.  Often, many constraints exist, having to do with costs, errors, time, and stability, which then define the aim to achieve a good degree of optimality.  There is a multitude of methods and approaches; see {\it e.g.} \cite{control_theory} for a general physics discussion.  From the latter perspective, control theory indeed combines statistical thermodynamics with the theory of dynamical systems.\\
Control is typically nonconservative.  The control or stimulus (such as the input current) gives rise to the dissipation of the used energy.  Moreover, in a more direct sense, the control can itself use nonconservative forces which makes it a driving scheme, and subject of nonlinear and nonequilibrium response studies.  The nonlinearity is present in applied feedback mechanisms but appears also from the very nature of the ``control solution'' which amounts to an inversion of the required response.  There is no guarantee that such an inversion works; in fact, it may not be possible.  In the present paper, it is accomplished via a phase shift.  In control theory,  ``differential flatness'' is often used as a condition for making the inversion possible, \cite{astrom}.  We have not studied how that flatness condition relates to our solution with the phase shift.\\

The present paper differs from standard control algorithms in a number of ways.  First of all, we employ a stochastic process as stimulus.  A Markov jump process is coupled to the probe.  We do use nonconservative driving in the jumps but not as control.  Our scheme is only {\it enabled} by the driving but the steering wheel is frenetic; {\it i.e., tunes} time-symmetric reactivities.\\
Frenesy is a quantity that is complementary to entropy.  It captures the idea of a time-symmetric dynamical activity and in that sense contrasts with fluxes or currents that have a (time-) direction.  In the case of Markov jump processes, as we use in the present paper, the frenesy is made from the escape rates and the time-symmetric part of the reactivities.  They are controlled by the symmetric parameters $a(x, \{\eta, \eta'\})$ in \eqref{kk}.  As mentioned in the introduction, it has been observed before that population and current selection can be achieved by their modulation when the nonequilibrium amplitude is high enough; see \cite{Maes_2016,non_dissipative_effects,fren,Basu_2015}. The present paper extends that idea to steering, which is a selection mechanism on the level of trajectories (for a probe).\\

One can ask here for (experimental) realizations. We believe that the presented program naturally leads to a construction plan for bio(mimetic)-engines.  After all, locomotion, contraction or biological work more generally, are produced via the coupling of probes ({\it e.g.} flagella) to nonequilibrium (sub)systems ({\it e.g.} molecular motors).  That is exactly our scheme where the driven Markov jump process, representing agitated molecules, is coupled to a much slower passive probe. 
A recent example of such a construction can be followed in \cite{faezeh}.  

Not all biological steering is frenetic but the novelty of our idea is to at least complement the use of external field protocols and of the tuned driving over energy landscapes with the possibility of changing the occupation statistics and hence the mean force via modulation of time-symmetric quantities.

\section{Conclusion}\label{sion}
Frenetic steering refers to the manipulation of parameters appearing in time-symmetric activity functions.  It is applicable for a slow probe or collective variable coupled to a nonequilibrium medium  where it adds the ability to manufacture forces.  As a matter of fact, the parameterization can be done via a phase shift variable, which creates a twist between the variable entropy fluxes and the excess dynamical activity or frenesy.\\
Various algorithms are possible, and we have selected just one that works well. We have explained the general principle, and how to make it work in a multidimensional setting. We have illustrated how to guide a probe to {\it simulate} the van der Pol cycle and Lorenz butterfly.  All that is obviously impossible for an equilibrium environment where the induced mean force on the system variables is always the gradient of a potential; guiding a probe or collective variable along a given trajectory requires coupling it with a nonequilibrium medium.\\

The idea may have more applications but also points to extra flexibility in  active or living processes: controlling the time-symmetric fluctuation sector is  a viable way to steer toward greater fitness, away from damage or even extinction.  No need to update the landscape of free energies, but for a given driving or fuelling, the creation or modifications of phase shifts between entropy fluxes and frenesy may adapt functionalities to a changing environment.

\section{Data availability}
Data are freely and directly available upon simple request from the authors.

\appendix\label{app}

\section{Shift and amplitude dependence}\label{appendix:cos_phi_dependence}
Consider the situation of a probe on the unit circle coupled to a nonequilibrium medium, whose state is denoted by $\eta$, with a small coupling parameter $\lambda$. Suppose we have a dynamics for the medium with a potential $U_\lambda(x, \eta) = U_0(\eta) + \lambda U_I(x, \eta)$ where $U_I(x, \eta)=h(\eta)\sin x$, a nonequilibrium driving $W(\eta_i, \eta_j)$ between every two states $\eta_i, \eta_j$ that is independent of $x$ and where the activity parameters are 
\begin{equation}
a_\lambda(x, \{\eta_i, \eta_j\})=a_0(\{\eta_i,\eta_j\})+\lambda b(\{\eta_i, \eta_j\})\cos(x+\varphi)
\end{equation}
where $\varphi$ is the same for every transition. 
The rates are then 
\begin{equation}
k(x, \eta_i, \eta_j) = a_\lambda(x, \{\eta_i, \eta_j\})\exp\big(\frac{\beta}{2}(U_\lambda(x, \eta_i)-U_\lambda(x, \eta_j) + W(\eta_i, \eta_j))\big)
\end{equation}
We view the dynamics as a perturbation with respect to the dynamics at $\lambda=0$. 
We show that to second order in $\lambda$: $\oint f(x)\id x\propto\cos \varphi$. We use the result from \cite{nongrad} that the nongradient part of the force can be written as 
\begin{equation}
f^\text{neq}(x)=-\frac{1}{\beta}\langle(D^\lambda-D^0)\nabla_x (S^\lambda-S^0)\rangle^0 + O(\lambda^3)
\label{eq:f_neq}
\end{equation}
The expectation is here over paths and we consider the system in its steady state again. $D$ is the frenesy and $S$ is the entropy flux, both functions of the path; see \cite{fren}. The superscript $\lambda$ signifies that the quantity is taken for the perturbed dynamics, and the superscript $0$ indicates the unperturbed dynamics. 
% To know what $D$ is we have to cover a bit of linear response theory. 
% We define the action $A(\omega)$ for a path $\omega$ as follows:
% \begin{equation}
% \frac{P^\lambda}{P^0}(\omega)=\exp(-A(\omega))
% \end{equation}
% and it can be written as follows:
% \begin{equation}
% A(\omega)
% \end{equation}
$D^\lambda - D^0$ depends on a path $\omega$ as 
\begin{align}
(D^\lambda -D^0)(\omega) = \int_0^t \id s\sum_{\eta\neq\eta(s)}\exp\big(\frac{\beta}{2}(U_0(\eta(s))-U_0(\eta) + W(\eta(s), \eta))\big)\\
\Big[\big(a(\{\eta(s), \eta\}) + \lambda b(\{\eta(s), \eta\})\cos(x+\varphi)\big)\exp\big(\lambda\frac{\beta}{2}(U_I(x, \eta(s))-U_I(x, \eta))\big)\notag\\ 
- a(\eta(s), \eta)\Big]-\sum_s\log\big(1+\lambda\frac{b(\{\eta(s^-), \eta(s)\})}{a(\{\eta(s^-), \eta(s)\})}\cos(x+\varphi)\big)\notag
\end{align}
where the sum over $s$ is a sum over the jump times. Upon expanding to first order in $\lambda$ and inserting the expression for $U_I(x,\eta)$, we get
\begin{align}
(D^\lambda -D^0)(\omega) = \lambda\int_0^t \id s\sum_{\eta\neq\eta(s)}\exp\big(\frac{\beta}{2}(U_0(\eta(s))-U_0(\eta) + W(\eta(s), \eta))\big)\notag\\
\Big[\big(a(\{\eta(s), \eta\})\frac{\beta}{2}(h(\eta(s))-h(\eta))\sin x + b(\{\eta(s), \eta\})\cos(x+\varphi)\big)\Big]\notag\\
-\lambda\sum_s\frac{b(\{\eta(s^-), \eta(s)\})}{a(\{\eta(s^-), \eta(s)\})}\cos(x+\varphi)\notag
\end{align}
$(S^\lambda - S^0)(\omega)$ has a much simpler form, 
\begin{equation}
(S^\lambda - S^0)(\omega)=\lambda\beta(\sum_sh(\eta(s^-))-h(\eta(s)))\sin x=\lambda\beta(h(\eta(0))-h(\eta(t)))\sin x
\end{equation}
and so 
\begin{equation}
\nabla_x(S^\lambda - S^0)(\omega)=\lambda\beta(h(\eta(0))-h(\eta(t)))\cos x
\end{equation}
To calculate $\oint f^\text{neq}(x)\id x$. we switch the expectation operation and the integral over $x$. Then, for small $\lambda$, to order $\lambda^2$,
\begin{align}
-\frac{1}{\beta}\oint \id x(D^\lambda -D^0)(\omega)\nabla_x(S^\lambda-S^0)(\omega) = \\ 
-\lambda^2\int_0^t ds \oint \id x\sum_{\eta\neq\eta(s)}\exp\big(\frac{\beta}{2}(U_0(\eta(s))-U_0(\eta) + W(\eta(s), \eta))\big)\notag\\
\Big[a(\{\eta(s), \eta\})\frac{\beta}{2}(h(\eta(s))-h(\eta))\sin x + b(\{\eta(s), \eta\})\cos(x+\varphi)\Big]\notag\\
(h(\eta(0))-h(\eta(t)))\cos x\notag\\
+\lambda^2\oint \id x\sum_s\frac{b(\{\eta(s^-), \eta(s)\})}{a(\{\eta(s^-), \eta(s)\})}\cos(x+\varphi)(h(\eta(0))-h(\eta(t)))\cos x\notag
\end{align}
Using the sum formul{\ae} for sine and cosine,
%now that $\sin x\cos x=1/2\sin(2x)$, $\cos(x+\varphi)=\cos x\cos \varphi-\sin x\sin(\varphi)$ and $\cos^2(x)=(1+\cos(2x))/2$, we get:
\begin{align}
-\frac{1}{\beta}\oint \id x(D^\lambda -D^0)(\omega)\nabla_x(S^\lambda-S^0)(\omega) = \\ 
-\lambda^2\,\pi\,\cos \varphi\int_0^t \id s\sum_{\eta\neq\eta(s)}\exp\big(\frac{\beta}{2}(U_0(\eta(s))-U_0(\eta) + W(\eta(s), \eta))\big)\notag\\
b(\{\eta(s), \eta\})(h(\eta(0))-h(\eta(t)))\notag\\
+\lambda^2\pi\,\cos \varphi\sum_s\frac{b(\{\eta(s^-), \eta(s)\})}{a(\{\eta(s^-), \eta(s)\})}(h(\eta(0))-h(\eta(t)))\notag
\end{align}
to order $\lambda^2$.
Since that is true for every path $\omega$, it shows that $\oint f^\text{neq}(x)\id x\propto\cos \varphi$ to second order in $\lambda$. Furthermore, from this formula, we also see that multiplying $b$ with the same factor $d$ for all transitions, $\oint f^\text{neq}(x)\,\id x$ to second order in $\lambda$ is multiplied by $d$ as well.

\section{Reduction of the multi-dimensional case to one dimension}\label{dim}
\label{appendix:multiple_dimensions}
Suppose that the probe lives on a higher-dimensional torus, $x\in T^n$. 
%We can extend the simple system for which some calculations were made in Appendix \ref{appendix:cos_phi_dependence} to multiple dimensions. The calculations there give an idea for how to create a force for which for every dimension we can choose the rotational part. 
The idea is to perturb the reactivities and the energy in an additive way over the different dimensions. We can freely choose the rotational part for every dimension. We show it here for $n=2$ dimensions, but it generalizes to an arbitrary number of dimensions.\\ 
Let us take the potential 
\begin{equation}
U_\lambda(x, y, \eta) = U_0(\eta) + \lambda h_x(\eta)\sin x + \lambda h_y(\eta)\sin y
\end{equation}
and activity parameters,
\begin{equation}
a_\lambda(x,y,\{\eta_i, \eta_j\}) = a_0(\{\eta_i, \eta_j\}) + \lambda b_x(\{\eta_i, \eta_j\})\cos(x + \varphi_x) + \lambda b_y(\{\eta_i, \eta_j\})\cos(y + \varphi_y)
\end{equation}
and also again a nonequilibrium driving $W(\eta_i, \eta_j)$ between every two states $\eta_i, \eta_j$ that is independent of $x$ and $y$.
We use \eqref{eq:f_neq} again, now to calculate $\oint f_x(x,y)\id x$. The case $\oint f_y(x,y)\,\id y$ is completely analogous. Now, for $(D^\lambda-D^0)(\omega)$ to first order in $\lambda$:
\begin{align}
(D^\lambda -D^0)(\omega) = \lambda\int_0^t \id s\sum_{\eta\neq\eta(s)}\exp\big(\frac{\beta}{2}(U_0(\eta(s))-U_0(\eta) + W(\eta(s), \eta))\big)\\
\Big[\big(a(\{\eta(s), \eta\})\frac{\beta}{2}\big((h_x(\eta(s))-h_x(\eta))\sin x + (h_y(\eta(s))-h_y(\eta))\sin y\big) +\notag \\
b_x(\{\eta(s), \eta\})\cos(x+\varphi_x) + b_y(\{\eta(s), \eta\})\cos(y+\varphi_y)\Big]\notag\\
-\lambda\sum_s\frac{b_x(\{\eta(s^-), \eta(s)\})}{a(\{\eta(s^-), \eta(s)\})}\cos(x+\varphi_x) - \lambda\sum_s\frac{b_y(\{\eta(s^-), \eta(s)\})}{a(\{\eta(s^-), \eta(s)\})}\cos(y+\varphi_y)\notag
\end{align}
and %for $\partial(S^\lambda-S^0)(\omega)/\partial x$ and $\partial(S^\lambda-S^0)(\omega)/\partial y$:
\begin{align}
\partial(S^\lambda-S^0)(\omega)/\partial x & = \lambda\beta(h_x(\eta(0))-h_x(\eta(t)))\cos x\\
\partial(S^\lambda-S^0)(\omega)/\partial y & = \lambda\beta(h_y(\eta(0))-h_y(\eta(t)))\cos y\notag
\end{align}
We calculate $\oint f_x^\text{neq}(x,y)\id x$; the result for the other dimensions is analogous. Let us again switch the expectation with the integral, and work to second order in $\lambda$,
\begin{align}
-\frac{1}{\beta}\oint \id x(D^\lambda -D^0)(\omega)\partial(S^\lambda-S^0)(\omega)/\partial x = \\ 
-\lambda^2\int_0^tds\oint\id x\sum_{\eta\neq\eta(s)}\exp\big(\frac{\beta}{2}(U_0(\eta(s))-U_0(\eta) + W(\eta(s), \eta))\big)\notag\\
\Big[a(\{\eta(s), \eta\})\frac{\beta}{2}\big((h_x(\eta(s))-h_x(\eta))\sin x + (h_y(\eta(s)) - h_y(\eta))\sin y\big)\notag\\
+b_x(\{\eta(s), \eta\})\cos(x+\varphi_x) + b_y(\{\eta(s), \eta\})\cos(y+\varphi_y)\Big](h_x(\eta(0))-h_x(\eta(t)))\cos x\notag\\
+\lambda^2\oint \id x\sum_s\frac{b_x(\{\eta(s^-), \eta(s)\})}{a(\{\eta(s^-), \eta(s)\})}\cos(x+\varphi_x)(h_x(\eta(0))-h_x(\eta(t)))\cos x\notag\\
+\lambda^2\oint \id x\sum_s\frac{b_y(\{\eta(s^-), \eta(s)\})}{a(\{\eta(s^-), \eta(s)\})}\cos(y+\varphi_y)(h_x(\eta(0))-h_x(\eta(t)))\cos x\notag
\end{align}
Using again the sum rules for sine and cosine,
%that $\sin x\cos x=1/2\sin(2x)$, $\cos(x+\varphi)=\cos x\cos \varphi-\sin x\sin(\varphi)$ and $\cos^2(x)=(1+\cos(2x))/2$ we get:
\begin{align}
-\frac{1}{\beta}\oint \id x(D^\lambda -D^0)(\omega)\partial(S^\lambda-S^0)(\omega)/\partial x = \\ 
-\lambda^2\cos(\varphi_x)\pi\int_0^t \id s \sum_{\eta\neq\eta(s)}\exp\big(\frac{\beta}{2}(U_0(\eta(s))-U_0(\eta) + W(\eta(s), \eta))\big)\notag\\
b_x(\{\eta(s), \eta\})(h_x(\eta(0))-h_x(\eta(t)))\notag\\
+\lambda^2\cos(\varphi_x)\pi\sum_s\frac{b_x(\{\eta(s^-), \eta(s)\})}{a(\{\eta(s^-), \eta(s)\})}(h_x(\eta(0))-h_x(\eta(t)))\notag
\end{align}
to order $\lambda^2$. All terms involving $y$ have disappeared and we can control $\oint f_x(x,y)\id x$ via $\varphi_x$, to second order in $\lambda$, as in the one-dimensional case, as if there were no perturbations in the second $y-$dimension. 
% Let us again switch the expectation operation and the integral in order to get an expression for $\oint f^\text{neq}(x)\id x$ and similarly for $y$, we see that for every dimension, when we take the integral over that dimension, the terms involving perturbations for the other dimensions disappear. 

\section{Using a four-state nonequilibrium medium}\label{app:4_states}
%\textcolor{olive-green}{(Misschien moet ik de figuren zij aan zij zetten met de figuren voor 3 toestanden zodat ze makkelijk te vergelijken zijn?) Nee, hier en zo is beter}

We can wonder whether the steering performance increases when including more states in the nonequilibrium medium, or when increasing the driving amplitude more.  The answer is that the forces that can be obtained remain of the same order, and that the dependence on the driving saturates as well.  Here we illustrate the case using a nonequilibrium medium with 4 instead of 3 states, described around \eqref{act3}.  We discuss similarities and some differences.\\ 

The nonequilibrium model with 4 states is a straightforward adaptation of the model with 3 states described around \eqref{act3} to 4 states. We describe the dynamics here. 
We now have $\eta\in\{0, 1, 2, 3\}$.  The transition rates $k_x(\eta,\eta')$ are still given by \eqref{kk}. There is a driving with constant magnitude $W(0, 1)=W(1, 2)=W(2, 3)=W(3, 0)=\varepsilon$. 
For the interaction energy we now have $U(x, \eta)=\lambda\sin(x-\eta\pi/2)$ and for the activity parameters:
\begin{equation}\label{act4}
a(x, \{\eta, \eta'\})=\left\{
\begin{array}{rl}
1+\lambda b\cos(x-\eta\pi/2+\varphi)&\text{if } \eta'-\eta\equiv 1\pmod{4}\\
1+\lambda b\cos(x-(\eta-1)\pi/2+\varphi)&\text{if } \eta'-\eta\equiv 3\pmod{4}
\end{array}\right.
\end{equation}
When $\eta'-\eta\equiv 2\pmod{4}$, $a(x, \{\eta, \eta'\})=0$ because the states are on a cycle.

\begin{figure}
    \centering
    \includegraphics[height=0.3\textheight]{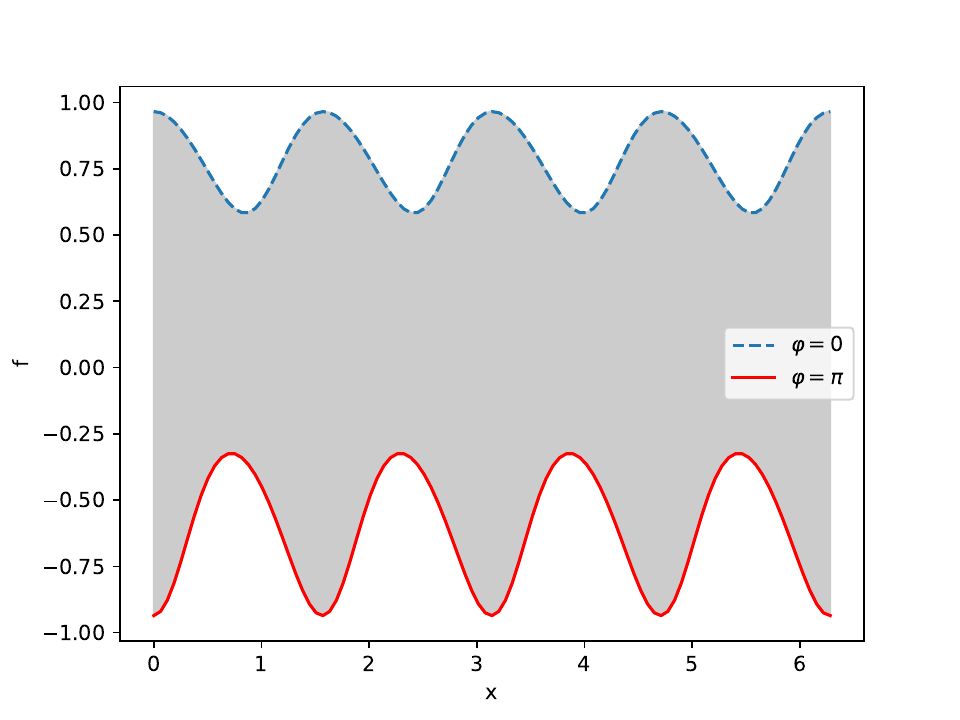}
    \caption{The force $f(x)$ for $b=0.99,\ \lambda=1,\ \beta=1\ \text{and }\varepsilon=5$, for $\varphi=0$ and $\varphi=\pi$ in the nonequilibrium model with 4 states. The shaded region indicates that any value of the force in that region is attainable by a suitable choice for $\varphi$.}
    \label{fig:f_2_phis_4_states}
\end{figure}

Fig.~\ref{fig:f_2_phis_4_states} shows the force for $\varphi= 0\text{ and }\varphi=\pi$ depending on the position $x$, using parameter values $b=0.99,\ \lambda=1,\ \beta=1\ \text{and }\varepsilon=5$. Indeed the curves have a smaller amplitude and a smaller period than in Fig.~\ref{fig:f_2_phis_3_states}. Here we have 4 instead of 3 peaks for both curves. 

\begin{figure}
    \centering
    \includegraphics[height=0.3\textheight]{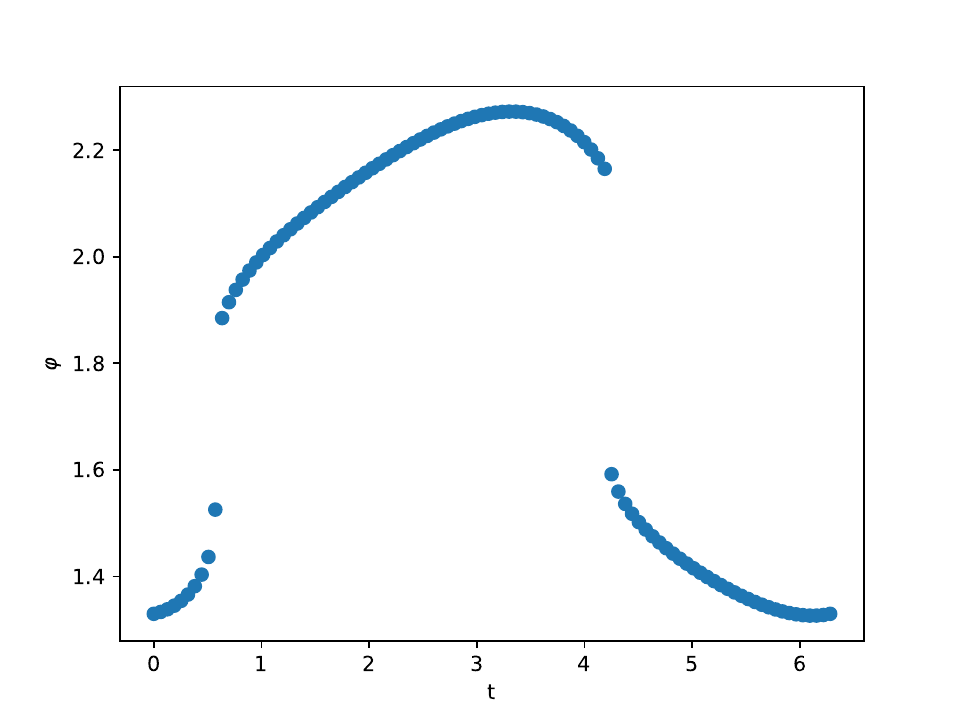}
    \caption{The shift $\varphi(t)$ to reprodude the trajectory $x(t)=0.1\sin t$, using 100 data points between $t=0$ and $t=2\pi$ and for parameter values $b=0.99,\ \lambda=1,\ \beta=1\ \text{and }\varepsilon=5$ for the nonequilibrium model with 4 states.}
    \label{fig:phi_t_4_states}
\end{figure}

Fig.~\ref{fig:phi_t_4_states} shows $\varphi(t)$ for the given trajectory $x(t)=0.1\,\sin t$ for 100 data points between $t=0$ and $t=2\pi$. Parameter values $b=0.99,\ \lambda=1,\ \beta=1\ \text{and }\varepsilon=5$ were used again. The two jumps in the value for $\varphi$ ({\it e.g.}, from around 1.6 to around 1.9) exist because $f(x(t), \varphi(t))=\dot{x}(t)$ has multiple solutions for $\varphi(t)$ for some values of $x(t)$ and $\dot{x}(t)$, and the algorithm can jump to the other solution (and in some cases has no other choice because for some values of $x(t)$ and $\dot{x}(t)$ there is only one solution for $\varphi(t)$). Aside from the fact that there are no jumps in the value for $\varphi(t)$, the curve in Fig.~\ref{fig:phi_t_4_states} has the same overall shape as the curve in Fig.~\ref{fig:phi_t_3_states}: they are both concave and attain their maximal value at $t$ a little bit smaller than 4. This is undoubtedly due to the models being very similar. The dependence of the potential $U$ and the activity parameters $a$ on the slow macroscopic degrees of freedom, the states of the nonequilibrium medium and the phase shifts is very similar; the difference lies only in the number of `steps' there are.\\ 
%That means for example that for the same trajectory $x(t)$ with one rescaling there are no jumps in the values of the $\varphi(t)$ while in another rescaling there are jumps.
%Making a loop, there are no transitions when $\eta-\eta'\equiv 2\pmod{4}$.  
%The states of each spin are of course still arranged in a cycle, so there are no transitions for two states of a spin $\eta^x_i, \eta^x_j$ when $\eta^x_i-\eta^x_j\equiv 2\pmod{4}$ and similarly for $\eta^y$. 
%The values for $\varphi$ are roughly in the range $[1.3, 1.6]\cup [1.9, 2.2]$ so there is a gap between around 1.6 to around 1.9. That is similar to what happens in Fig.~\ref{fig:phi_t_4_states}. 

Let us now move to the illustrations in Section \ref{sec:illustrations}.\\
The first thing to say is that there is no change in the actual trajectories, i.e., Fig.~\ref{fig:van_der_pol_steering_3_states} and Fig.~\ref{fig:lorenz_steering_3_states} do not change when using 4 states in the nonequilibrium medium. The point is that the grid used to approximate the dynamics is exactly the same. It does not matter that the time-derivative of the slow macroscopic variable is determined by the 3-state or the 4-state model for the medium. For all grid points the result is exactly the same.\\

\begin{figure}
    \centering
    \includegraphics[height=0.3\textheight]{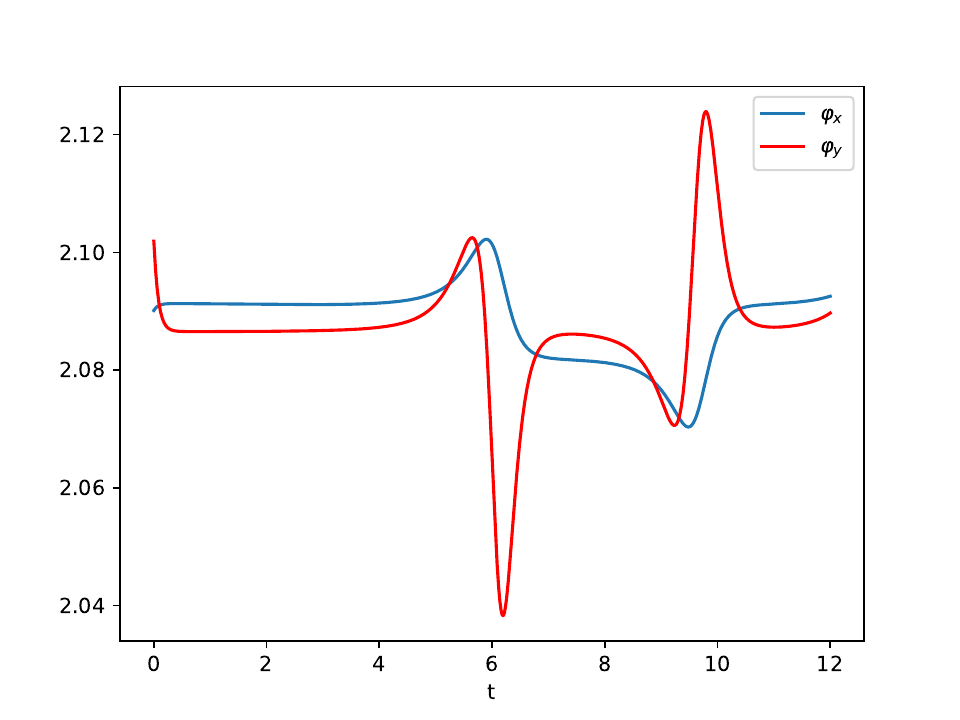}
    \caption{Phase shifts $\varphi_x(t)$ and $\varphi_y(t)$ for the medium dynamics coupled to the probe that follows the van der Pol oscillator, for $\mu=1.5$, initial state $x=3, y=0$, time step $0.01$ and $T=12$ for the nonequilibrium model with 4 states.}
    \label{fig:van_der_pol_phi_t_4_states}
\end{figure}

What is different is the required choice of phase shifts. For the van der Pol oscillator and for the same values for the parameters, the phase shifts $\varphi_x(t)$ and $\varphi_y(t)$ are shown in Fig.~\ref{fig:van_der_pol_phi_t_4_states}. Both curves again have a similar shape as in Fig.~\ref{fig:van_der_pol_phi_t_3_states} for the model with 3 states.

\begin{figure}
    \centering
    \subfigure[\,$\varphi_x(t)$]{
        \includegraphics[height=0.25\textheight]{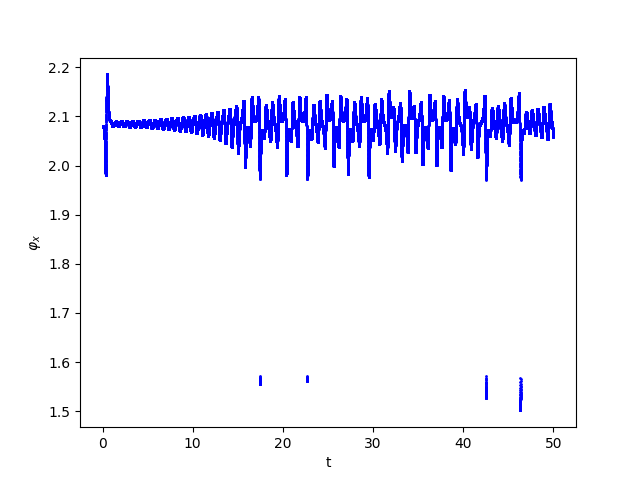}
        \label{fig:lorenz_phi_x_t_4_states}
    }
    \subfigure[\,$\varphi_z(t)$]{
        \includegraphics[height=0.25\textheight]{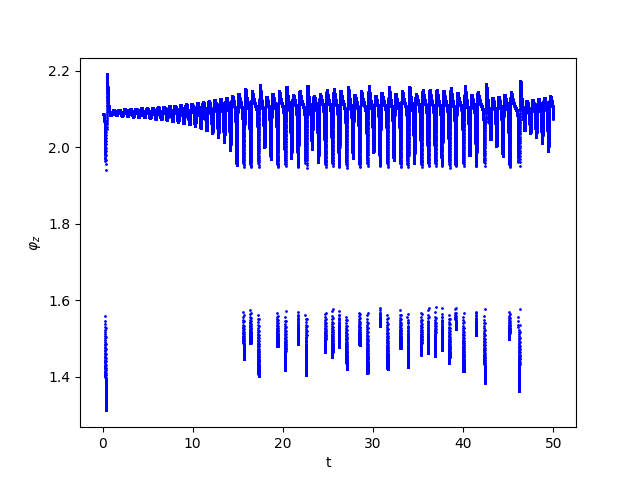}
        \label{fig:lorenz_phi_z_t_4_states}
    }
    \caption{Shift $\varphi_x(t) \text{ and } \varphi_z(t) $ required for simulating the Lorenz model, for initial state $x=0, y=1, z=0$, time step $0.001$ and $T=50$ using the nonequilibrium medium, with 4 states.}
    \label{fig:lorenz_phi_t_4_states}
\end{figure}

For the Lorenz dynamics, Fig.~\ref{fig:lorenz_phi_t_4_states} shows the shifts $\varphi_x(t)$ and $\varphi_z(t)$ for a trajectory using the same parameter values as in Fig.~\ref{fig:lorenz_phi_t_3_states}. There are again similarities with the plots shown in Fig.~\ref{fig:lorenz_phi_t_3_states}, but, for the 4 state model, there are jumps in the values for $\varphi_x(t)$ and $\varphi_z(t)$, like we had in Fig.~\ref{fig:phi_t_4_states}. The values for $\varphi$ are roughly in the range $[1.3, 1.6]\cup [1.9, 2.2]$ so there is a gap between around 1.6 to around 1.9.

\section{Scaling}\label{appendix:scaling}
The force that can be generated by the four-state cycle system described in Section \ref{sec:steering_the_force} is bounded. Consider again the dynamical system $\dot{x}=g(x)$.
If we can find an interval $[x_0, x_1]$ for which $g(x)$ is bounded, then we can do a rescaling of the system so that for the rescaled system the required force always falls between the two curves in Fig.~\ref{fig:f_2_phis_3_states}. We rescale in the following way: $x=Ax'$ for some $A>1$, big enough. A good choice for $A$ would be $\max \big\{|\dot{x}|\,\big|x\in [x_0, x_1]\big\}/0.2$ (notice in Fig.~\ref{fig:f_2_phis_3_states} that for the curve for $\varphi=0$: $f(x)>0.2$ and that for the curve for $\varphi=\pi$: $f(x)<-0.2$). 
The rescaled system obeys
\begin{equation}
\dot{x}'=g'(x')=\dot{x}/A=g(x)/A    
\end{equation}
For $x\in[x_0, x_1]$, we can then find $\varphi'(x')$ (we write $\varphi'$ instead of $\varphi$ because it is a function of $x'$), so that the force generated by the medium is equal to the force required by the rescaled dynamical system: $f(x', \varphi'(x'))=g'(x')$. We put $\varphi(x)=\varphi'(x')$, and given $\varphi'(x')$ we can for any $x\in[x_0, x_1]$ obtain $g(x)=Af(x', \varphi'(x'))$.  This method can be straightforwardly generalized to multiple dimensions.

\pagebreak

\bibliographystyle{hunsrt}  %ieeetr
\bibliography{frenstepaper}

\begin{thebibliography}{10}

\bibitem{islamic_gardens}
D.~Fairchild Ruggles.
\newblock {\em Islamic gardens and landscapes}.
\newblock Penn studies in landscape architecture. University of Pennsylvania Press, Philadelphia, (2008).

\bibitem{cauchy}
Augustin-Louis Cauchy.
\newblock Methode generale pour la resolution des systemes d'equations simultanees.
\newblock {\em Comptes Rendus Hebd. Séances Acad. Sci. Paris}, 25:536--538, (1847).

\bibitem{timematters}
Roman {Belousov}, Sabrina {Savino}, Prachiti {Moghe}, Takashi {Hiiragi}, Lamberto {Rondoni}, and Anna {Erzberger}.
\newblock When time matters: Poissonian cellular potts models reveal nonequilibrium kinetics of cell sorting.
\newblock {\em arXiv e-prints}, (2023), 2306.04443.

\bibitem{non_dissipative_effects}
Christian Maes.
\newblock {\em Non-Dissipative Effects in Nonequilibrium Systems}.
\newblock Springer International Publishing AG, Cham, (2017).

\bibitem{frensteering}
Bram Lefebvre and Christian Maes.
\newblock Frenetic steering in a nonequilibrium graph.
\newblock {\em Journal of Statistical Physics}, \textbf{190}(4):90, (2023).

\bibitem{kinetic_uncertainty_relation}
Ivan~Di Terlizzi and Marco Baiesi.
\newblock Kinetic uncertainty relation.
\newblock {\em J. Phys. A: Math. Theor.}, \textbf{52}(2):02LT03, (2019).

\bibitem{inadequacy_of_entropy}
Rolf Landauer.
\newblock Inadequacy of entropy and entropy derivatives in characterizing the steady state.
\newblock {\em Physical review. A, General physics}, \textbf{12}(2):636--638, (1975).

\bibitem{hb}
Christian Maes and Karel Netočný.
\newblock Heat bounds and the blowtorch theorem.
\newblock {\em Ann.~H.~Poincaré}, \textbf{14}(5):1193--1202, (2013).

\bibitem{kinetic_proofreading}
John Hopfield.
\newblock Kinetic proofreading: A new mechanism for reducing errors in biosynthetic processes requiring high specificity.
\newblock {\em Proc. Natl. Acad. Sci. U.S.A.}, \textbf{71}(10):4135--4139, (1974).

\bibitem{nongrad}
Christian Maes and Karel Netočný.
\newblock {Nonequilibrium corrections to gradient flow}.
\newblock {\em Chaos: An Interdisciplinary Journal of Nonlinear Science}, \textbf{29}(7):073109, (2019).

\bibitem{fren}
Christian Maes.
\newblock Frenesy: Time-symmetric dynamical activity in nonequilibria.
\newblock {\em Physics Reports}, \textbf{850}:1--33, (2020).

\bibitem{control_theory}
John Bechhoefer.
\newblock {\em Control Theory for Physicists}.
\newblock Cambridge University Press, Cambridge, (2021).

\bibitem{astrom}
Karl~Johan. Åstrom and Richard~M.. Murray.
\newblock {\em Feedback systems: an introduction for scientists and engineers}.
\newblock Princeton University Press, Princeton, second edition, (2021).

\bibitem{python}
Guido Van~Rossum and Fred~L. Drake.
\newblock {\em Python 3 Reference Manual}.
\newblock CreateSpace, Scotts Valley, CA, (2009).

\bibitem{numpy}
Charles~R. Harris, K.~Jarrod Millman, St{\'{e}}fan~J. van~der Walt, Ralf Gommers, Pauli Virtanen, David Cournapeau, Eric Wieser, Julian Taylor, Sebastian Berg, Nathaniel~J. Smith, Robert Kern, Matti Picus, Stephan Hoyer, Marten~H. van Kerkwijk, Matthew Brett, Allan Haldane, Jaime~Fern{\'{a}}ndez del R{\'{i}}o, Mark Wiebe, Pearu Peterson, Pierre G{\'{e}}rard-Marchant, Kevin Sheppard, Tyler Reddy, Warren Weckesser, Hameer Abbasi, Christoph Gohlke, and Travis~E. Oliphant.
\newblock Array programming with {NumPy}.
\newblock {\em Nature}, 585(7825):357--362, (2020).

\bibitem{matplotlib}
J.~D. Hunter.
\newblock Matplotlib: A 2d graphics environment.
\newblock {\em Computing in Science \& Engineering}, 9(3):90--95, (2007).

\bibitem{scipy}
Pauli Virtanen, Ralf Gommers, Travis~E. Oliphant, Matt Haberland, Tyler Reddy, David Cournapeau, Evgeni Burovski, Pearu Peterson, Warren Weckesser, Jonathan Bright, St{\'e}fan~J. {van der Walt}, Matthew Brett, Joshua Wilson, K.~Jarrod Millman, Nikolay Mayorov, Andrew R.~J. Nelson, Eric Jones, Robert Kern, Eric Larson, C~J Carey, {\.I}lhan Polat, Yu~Feng, Eric~W. Moore, Jake {VanderPlas}, Denis Laxalde, Josef Perktold, Robert Cimrman, Ian Henriksen, E.~A. Quintero, Charles~R. Harris, Anne~M. Archibald, Ant{\^o}nio~H. Ribeiro, Fabian Pedregosa, Paul {van Mulbregt}, and {SciPy 1.0 Contributors}.
\newblock {{SciPy} 1.0: Fundamental Algorithms for Scientific Computing in Python}.
\newblock {\em Nature Methods}, 17:261--272, (2020).

\bibitem{strogatz}
Steven Strogatz.
\newblock {\em Nonlinear Dynamics and Chaos: With Applications to Physics, Biology, Chemistry and Engineering}.
\newblock Westview Press, Cambridge, 1st paperback print. edition, (2000).

\bibitem{vanderpol}
Balthasar van~der Pol.
\newblock On “relaxation-oscillations”.
\newblock {\em The London, Edinburgh, and Dublin Philosophical Magazine and Journal of Science}, \textbf{2}(11):978--992, (1926).

\bibitem{vanderpolmark}
Balthasar van~der Pol and Jan van~der Mark.
\newblock The heartbeat considered as a relaxation oscillation, and an electrical model of the heart.
\newblock {\em Philosophical Magazine}, \textbf{6}(38):763--775, (1928).

\bibitem{lorenz}
Edward Lorenz.
\newblock Deterministic nonperiodic flow.
\newblock {\em Journal of Atmospheric Sciences}, \textbf{20}(2):130--148, (1963).

\bibitem{saltzman}
Barry Saltzman.
\newblock Finite amplitude free convection as an initial value problem—i.
\newblock {\em Journal of Atmospheric Sciences}, \textbf{19}(4):329 -- 341, (1962).

\bibitem{hilborn}
Robert Hilborn.
\newblock {\em Chaos and Nonlinear Dynamics: An Introduction for Scientists and Engineers}.
\newblock Oxford University Press, New York, 2nd ed.; repr. edition, (2003).

\bibitem{berge}
Pierre Bergé, Yves Pommeau, and Christian Vidal.
\newblock {\em Order within Chaos: Towards a Deterministic Approach to Turbulence}.
\newblock John Wiley \& Sons, New York, (1984).

\bibitem{shen}
Bo-Wen Shen.
\newblock Nonlinear feedback in a six-dimensional lorenz model: impact of an additional heating term.
\newblock {\em Nonlinear processes in geophysics}, \textbf{22}(6):749--764, (2015).

\bibitem{warwick}
Warwick Tucker.
\newblock A rigorous ode solver and smale’s 14th problem.
\newblock {\em Foundations of computational mathematics}, \textbf{2}(1):53--117, (2002).

\bibitem{Maes_2016}
Christian Maes.
\newblock What decides the direction of a current?
\newblock {\em Mathematics and Mechanics of Complex Systems}, 4(3–4):275–295, (2016).

\bibitem{Basu_2015}
Urna Basu and Christian Maes.
\newblock Nonequilibrium response and frenesy.
\newblock {\em Journal of Physics: Conference Series}, 638:012001, (2015).

\bibitem{faezeh}
Faezeh Khodabandehlou and Christian Maes.
\newblock Local detailed balance for active particle models.
\newblock (2024), 2401.11850.

\end{thebibliography}
\onecolumngrid

\end{document}